\pdfoutput=1

\documentclass[11pt,twoside,a4paper,cmspaper,final,collab]{cms-tdr}
\def\svnVersion{287181}\def\cmsCernNo{2013-037}\def\cmsCernDate{\today}\def\cmsMessage{Published in the Journal of High Energy Physics as \href{http://dx.doi.org/10.1007/JHEP04(2015)164}{\doi{10.1007/JHEP04(2015)164.}}}

\begin{document}\cmsNoteHeader{SMP-13-014}

\hyphenation{had-ron-i-za-tion}
\hyphenation{cal-or-i-me-ter}
\hyphenation{de-vices}
\RCS$Revision: 282684 $
\RCS$HeadURL: svn+ssh://svn.cern.ch/reps/tdr2/papers/SMP-13-014/trunk/SMP-13-014.tex $
\RCS$Id: SMP-13-014.tex 282684 2015-03-30 09:08:40Z hindrich $
\newlength\cmsFigWidth
\ifthenelse{\boolean{cms@external}}{\setlength\cmsFigWidth{0.85\columnwidth}}{\setlength\cmsFigWidth{0.4\textwidth}}
\ifthenelse{\boolean{cms@external}}{\providecommand{\cmsLeft}{top}}{\providecommand{\cmsLeft}{left}}
\ifthenelse{\boolean{cms@external}}{\providecommand{\cmsRight}{bottom}}{\providecommand{\cmsRight}{right}}
\newcommand{\FIG}[1]{Fig.~\ref{#1}\xspace}
\newcommand{\FORM}[1]{Eq.~\ref{#1}\xspace}
\newcommand{\TAB}[1]{Table~\ref{#1}\xspace}
\newcommand{\see}{\ensuremath{\sigma_{\eta \eta}}\xspace}
\newcommand{\nfp}{\ensuremath{I_{\gamma\mathrm{,nfp}}}\xspace}
\newcommand{\Ic}{\ensuremath{I_{\mathrm{c}}}\xspace}
\newcommand{\In}{\ensuremath{I_{\mathrm{n}}}\xspace}
\newcommand{\Ig}{\ensuremath{I_{\gamma}}\xspace}
\newcommand{\ptg}{\ensuremath{p_{\mathrm{T}}^\gamma}\xspace}
\newcommand{\ZZg}{\ensuremath{\Z\Z\gamma}\xspace}
\newcommand{\Zgg}{\ensuremath{\Z\gamma\gamma}\xspace}
\newcommand{\Zg}{\ensuremath{\Z\gamma}\xspace}
\newcommand{\SM}{SM\xspace}
\newcommand{\AGC}{aTGC\xspace}
\newcommand{\FSR}{FSR\xspace}
\providecommand{\fb}{\ensuremath{\,\text{fb}}\xspace}

\cmsNoteHeader{SMP-13-014}

\date{\today}

\title{Measurement of the \Zg production cross section in pp collisions at 8\TeV and search for anomalous triple gauge boson couplings}
\abstract{
The cross section for the production of \Zg in proton-proton collisions at 8\TeV is measured based on data collected by the CMS experiment at the LHC corresponding to an integrated luminosity of 19.5\fbinv. Events with an oppositely-charged pair of muons or electrons together with an isolated photon are selected. The differential cross section as a function of the photon transverse momentum is measured inclusively and exclusively, where the exclusive selection applies a veto on central jets. The observed cross sections are compatible with the expectations of next-to-next-to-leading-order quantum chromodynamics. Limits on anomalous triple gauge couplings of \ZZg and \Zgg are set that improve on previous experimental results obtained with the charged lepton decay modes of the \Z boson.
}

\hypersetup{%
pdfauthor={CMS Collaboration},%
pdftitle={Measurement of the Z gamma production cross section in pp collisions at 8 TeV and search for anomalous triple gauge boson couplings},%
pdfsubject={CMS},%
pdfkeywords={CMS, physics, aTGC}}

\maketitle

\section{Introduction}
The study of \Zg production in proton-proton~(pp) collisions at \TeVns{} energies represents an important test of the standard model~(\SM), which prohibits direct coupling between the \Z boson and the photon. Within the \SM \Zg production is primarily due to radiation of photons from initial-state quarks (ISR) or final-state leptons (FSR). However, new physics phenomena at higher energies may be manifested as an effective self-coupling among neutral gauge bosons, resulting in a deviation from their predicted zero values in the \SM. Models of anomalous triple gauge couplings (\AGC) have been introduced and discussed in Refs.~\cite{PhysRevD.30.1513, PhysRevD.57.2823, PhysRevD.47.4889}.

\begin{figure}[ht]
\centering
\includegraphics[width=0.3\textwidth]{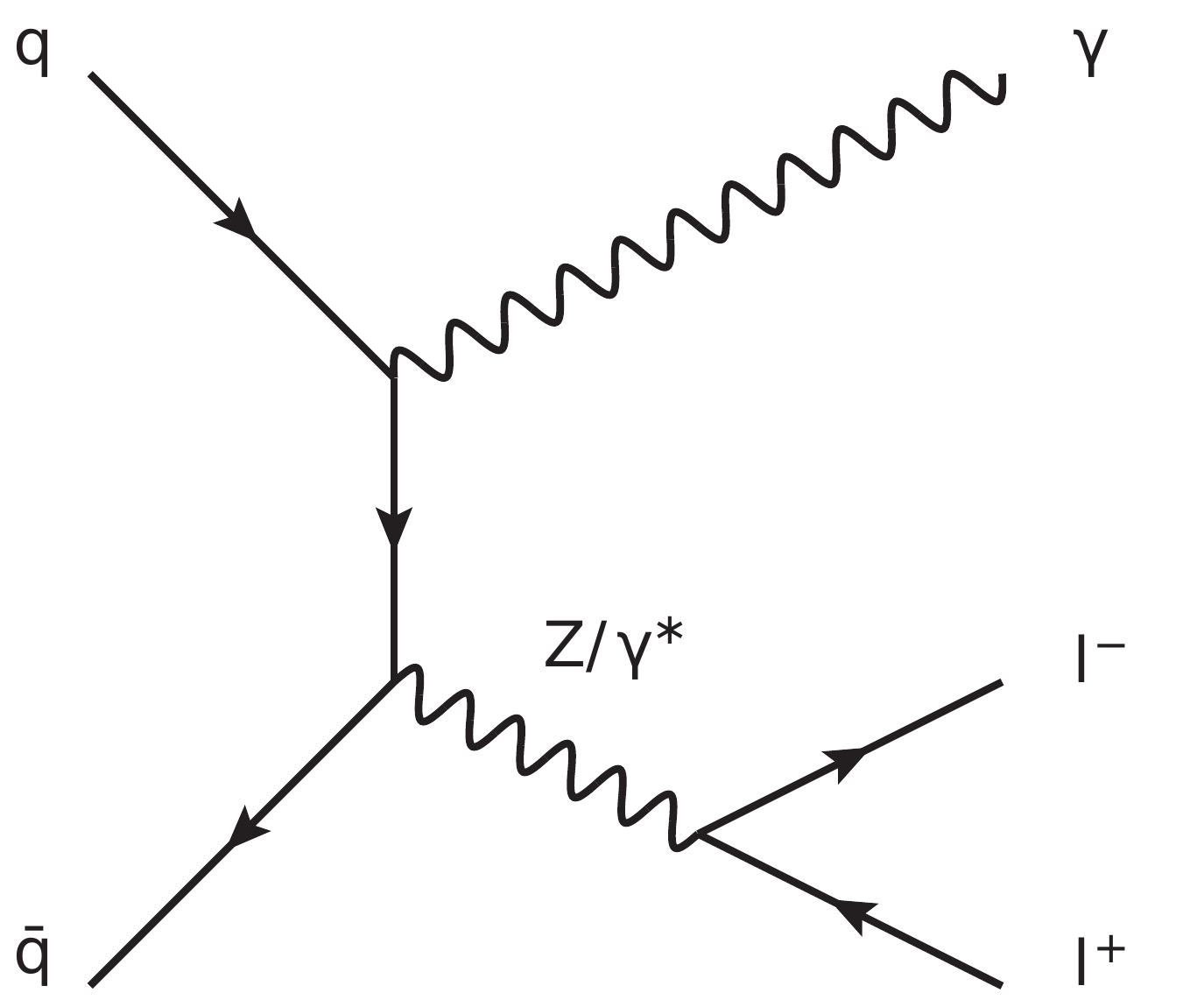}
\includegraphics[width=0.3\textwidth]{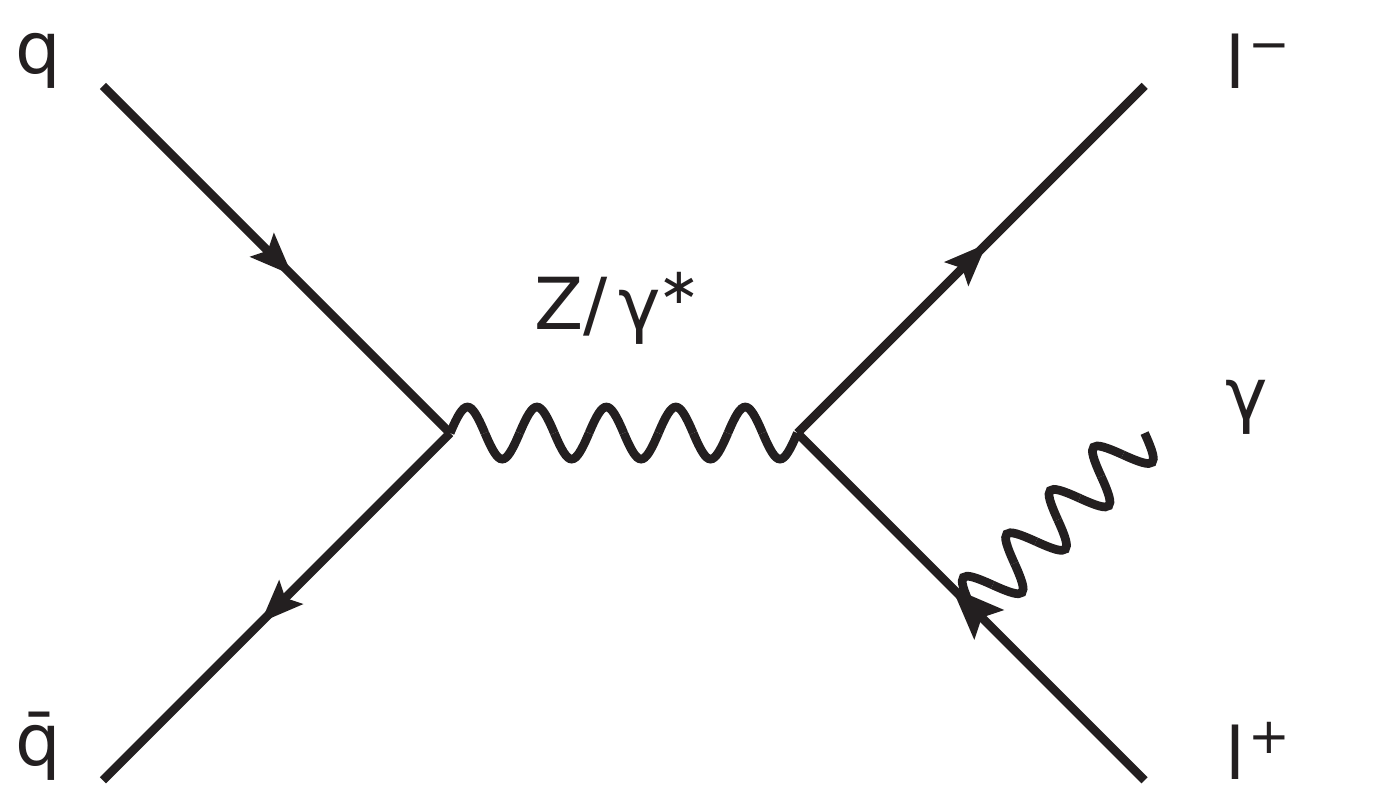}
\includegraphics[width=0.3\textwidth]{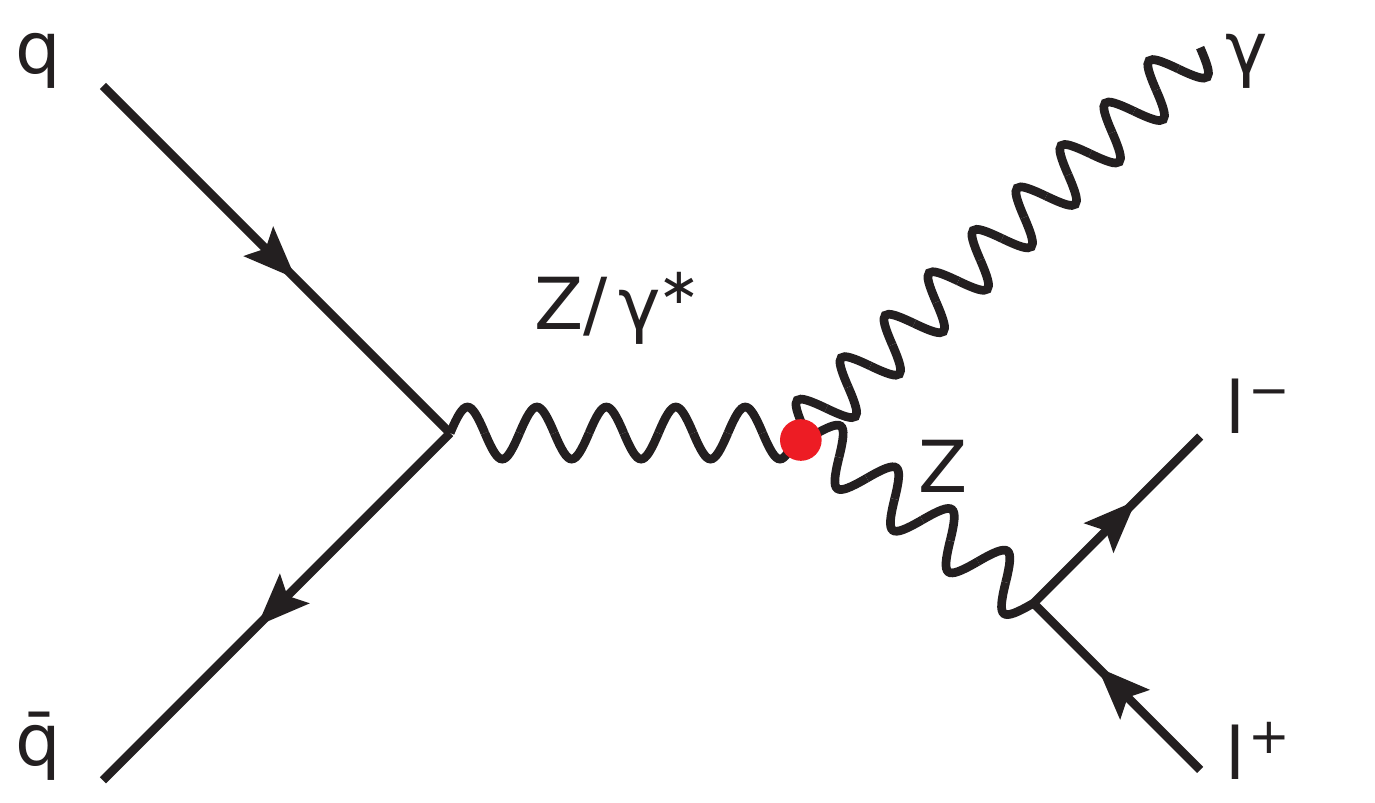}
\caption{Leading-order Feynman diagrams for \Zg production in pp collisions. Left: initial-state radiation. Center: final-state radiation. Right: diagram involving {\AGC}s that are forbidden in the \SM at tree level.}
\label{INTP1}
\end{figure}

This paper presents a measurement of \Zg production in pp collisions at a center-of-mass energy of 8\TeV, based on data collected with the CMS experiment in 2012, corresponding to an integrated luminosity of 19.5\fbinv. For this analysis the decays of the \Z boson into a pair of muons ($\mu^+\mu^-$) or electrons ($\Pep\Pem$) are considered. The processes of ISR and FSR contribute to $\ell^+\ell^-\gamma$ ($\ell=\mu,\Pe$) production in the \SM at leading order~(LO), and these are exemplified by the first two Feynman diagrams in \FIG{INTP1}. Photons can also be produced by jet fragmentation, but these photons are not considered as signal in the present analysis and are strongly suppressed by requiring that the photon is isolated. The production of \Zg through triple gauge couplings is represented by the third diagram in~\FIG{INTP1}.

Both ATLAS and CMS Collaborations have presented measurements of the inclusive \Zg cross section and searches for anomalous \ZZg and \Zgg couplings using data collected at a center-of-mass energy of 7\TeV~\cite{Aad:2013izg,PhysRevD.89.092005}. The larger 2012 data sample and the increased cross section at 8\TeV allow for the first measurement of the inclusive differential cross section for \Zg production as a function of the photon transverse momentum $\ptg$. Results on the differential \Zg cross section for events with no accompanying central jets, referred to as exclusive cross sections, are also presented, providing some insight into the effect of additional jets on the distribution of \ptg. The cross sections are measured for photons with $\ptg > 15$\GeV and restricted to a phase space defined by kinematic requirements on the final-state particles that are motivated by the experimental acceptance. In addition, the photon is required to be separated from the leptons by $\Delta R(\ell, \gamma) > 0.7$ where $\Delta R = \sqrt{\smash[b]{(\Delta\phi)^2 + (\Delta\eta)^2}}$, $\phi$ is the azimuthal angle and $\eta$ the pseudorapidity. Furthermore, the dilepton invariant mass is required to be above 50\GeV. With this selection the fraction of \FSR photons is reduced and photons originating from ISR or {\AGC}s are dominant.

\section{The CMS detector and particle reconstruction}
\label{sec:detector}

The central feature of the CMS apparatus is a superconducting solenoid of 6\unit{m} internal diameter, providing a magnetic field of 3.8\unit{T}. Within this superconducting solenoid volume are a silicon pixel and strip tracker, a lead tungstate crystal electromagnetic calorimeter~(ECAL), and a brass and scintillator hadron calorimeter~(HCAL),  each composed of a barrel and two endcap sections. Muons are measured in gas-ionization detectors embedded in the steel flux-return yoke outside the solenoid. Extensive forward calorimetry complements the coverage provided by the barrel and endcap detectors.

The silicon tracker measures charged particles within the pseudorapidity range $\abs{\eta}< 2.5$. The ECAL provides coverage in pseudorapidity $\abs{ \eta }< 1.479 $ in a barrel region (EB) and $1.479 <\abs{ \eta } < 3.0$ in two endcap regions (EE). A preshower detector consisting of two planes of silicon sensors interleaved with a total of three radiation lengths of lead is located in front of the EE regions. Muons are measured in the pseudorapidity range $\abs{\eta}< 2.4$, with detection planes made using three technologies: drift tubes, cathode strip chambers, and resistive-plate chambers.

The particle-flow (PF) algorithm~\cite{CMS-PAS-PFT-09-001,CMS-PAS-PFT-10-001,CMS-PAS-PFT-10-003} reconstructs and identifies each particle with an optimized combination of all subdetector information and categorizes reconstructed objects as photons, muons, electrons, charged hadrons, and neutral hadrons. The energy of photons is obtained from a cluster of energy depositions in ECAL crystals. The photon direction is determined by assuming it is associated to the main interaction vertex, defined as the vertex that has the highest value for the sum of $\pt^2$ of the associated tracks that is also compatible with the beam interaction point. The energy of electrons is determined from a combination of the track momentum at the main interaction vertex, the ECAL cluster energy, and the energy sum of all bremsstrahlung photons attached to the track~\cite{GSF_Electron_Reconstruction_CMS, Baffioni:2006cd}. The energy of muons is obtained from the corresponding track momentum measured in the silicon tracker and the muon detection system. The energy of charged hadrons is determined from a combination of the track momentum and the corresponding ECAL and HCAL energies. Finally, the energy of neutral hadrons is obtained from the corresponding ECAL and HCAL energies.

Jets used for the exclusive measurement presented in this paper are reconstructed from PF objects, clustered by the anti-\kt algorithm~\cite{Cacciari:2008gp, Cacciari:2011ma} with a distance parameter of 0.5. The measured jet momentum is the vectorial sum of all particle momenta in the jet and is found from the simulation to be within 5--10\%~\cite{Chatrchyan:2011ds} of the initial jet momentum over the whole \pt range and detector acceptance. An offset correction is applied to take into account the extra energy clustered in jets due to additional pp interactions within the same bunch crossing (pileup)~\cite{Cacciari:2008gn}.

A more detailed description of the CMS detector, together with a definition of the coordinate system used and the relevant kinematic variables, can be found in Ref.~\cite{Chatrchyan:2008zzk}.

\section{Signal and background modeling}
\label{sec:mcsamples}

Simulation samples for the signal process, $\ell^+\ell^-\gamma + n$ partons ($n = 0,1,2$) at matrix element level are produced with the event generator \SHERPA~1.4~\cite{Gleisberg:2008ta} for the muon and electron channels separately. The cross sections are calculated at next-to-leading order~(NLO) in quantum chromodynamics~(QCD) using \MCFM~6.4~\cite{Campbell:2012ft} and the CT10~\cite{Lai:2010vv} parton distribution functions~(PDF). Additional PDF sets are provided by CT10 to represent the uncertainties in the PDFs. These are used to estimate the PDF uncertainties in the cross sections following the prescription in Ref.~\cite{pdfunc}. The effect of using the CT10 PDF sets, where the value of the strong coupling constant $\alpha_s$ is varied in the fit, has been studied and is considered as an additional uncertainty. The uncertainties from factorization, renormalization, and photon fragmentation scales are estimated by varying each of these scales up and down by a factor of two. The uncertainty in the exclusive cross section calculation is obtained by following the recommendation in Ref.~\cite{Stewart:arXiv1107.2117} of adding in quadrature the scale uncertainties of the $\ell^+\ell^-\gamma$ NLO and the $\ell^+\ell^-\gamma + 1$ parton LO calculations. We also compare the measurement with a next-to-next-to-leading-order~(NNLO) calculation of the inclusive cross section provided by Ref.~\cite{Grazzini:arXiv1309.7000}.

The major sources of background to the \Zg process are Drell--Yan (DY) + jets, WW, WZ, ZZ, and $\ttbar$ production. These are simulated with the \MADGRAPH~4~\cite{Alwall:2007st} matrix element generator, using the CTEQ6L PDF set~\cite{Pumplin:2002vw}, and interfaced with \PYTHIA~6.4.26~\cite{Sjostrand:2006za} to describe parton showers, fragmentation, and initial and final state radiation of photons. The cross sections for $\ttbar$ and diboson production are normalized to the NLO QCD calculation from \MCFM. The DY+jets sample is normalized to the NNLO~QCD calculation of  \FEWZ~\cite{journals/cphysics/GavinLPQ11}. It is used to describe the background of nonprompt and misidentified photons. All events containing a prompt photon that passes the signal requirements are removed from this sample. The QCD simulation, which is used for the background determination, is produced using \PYTHIA. All samples are passed through a detailed simulation of the CMS detector based on \GEANTfour~\cite{geant4} and reconstructed using the same algorithms as used for data.

\section{Event selection}
\label{SEL}
The measurements presented in this paper rely on the reconstruction and identification of isolated muons, electrons, and photons. The exclusive cross section measurement is also dependent on the reconstruction of jets. Details of the identification and selection of muons (electrons) can be found in Ref.~\cite{Chatrchyan:2012xi} (Ref.~\cite{CMS-DP-2013-003}).

Leptons from \Z boson decays are typically isolated, \ie, separated in $\Delta R$ from other particles. A requirement on the lepton isolation is used to reject leptons produced in decays of hadrons. The muon isolation is based on tracks from the main interaction vertex as this is always identified as the source of the lepton pair. The isolation variable $I_{\text{trk}}$ is defined as the \pt sum of all tracks except for the muon track originating from the main interaction vertex within a cone of $\Delta R(\mu, \mathrm{track}) < 0.3$. The value of $I_{\text{trk}}$ is required to be less than 10\% of the muon \pt. For electrons the isolation variable is the \pt sum of neutral hadrons, charged hadrons, and photon-like PF objects in a cone of $\Delta R < 0.3$ around the electron. Contributions of the electron to the isolation variables are suppressed using a veto region. This isolation variable is required to be smaller than 10\% (15\%) of the electron \pt for electrons in the EB (EE). An event-by-event correction is applied, which keeps the selection efficiency constant as a function of pileup interactions~\cite{Cacciari:2007}.

Quality selection criteria are applied to the reconstructed photons to suppress the background from hadrons misidentified as photons. The ratio of the energy deposition in the HCAL tower behind the ECAL cluster to the energy deposition in the ECAL has to be below 5\%. The background is further suppressed by a requirement on the shower shape variable $\see$~\cite{sieie1}, which measures the shower width along the $\eta$ direction in a 5$\times$5 matrix of crystals centered on the crystal of highest energy in the cluster.
The electromagnetic shower produced by a photon is expected to have a small value of $\see$. Therefore, a selection of $\see < 0.011$ ($0.033$) in the EB (EE) region is applied. To suppress electrons misidentified as photons, photon candidates are rejected if measurements in the silicon pixel detector are found and these measurements are consistent with an electron, which is incident on the ECAL at the location of the photon cluster. The isolation variables based on PF charged hadrons $\Ic$, neutral hadrons $\In$, and photon objects $\Ig$ are calculated as the sum of \pt in a cone of $\Delta R < 0.3$ around the photon. Contributions of the photon itself are suppressed using a veto region. Again, the isolation values are corrected for the average energy deposition due to pileup. The isolation requirements used in the EB region are $\In < 1.0 + 0.04\ptg$, $\Ig < 0.7 + 0.005\ptg$, and $\Ic < 1.5$. In the EE region $\In < 1.5 + 0.04\ptg$, $\Ig < 1.0 + 0.005\ptg$, and $\Ic < 1.2$ are required.

Events are selected online using a dimuon or dielectron trigger with thresholds of $\pt > 17$\GeV for the leading and $\pt > 8$\GeV for the subleading lepton. Candidate events are required to have two same-flavor opposite-charge selected leptons and a selected photon. Muons with $\abs{\eta} < 2.4$ relative to the main interaction vertex are selected, while electrons need to satisfy $\abs{\eta_\mathrm{SC}}< 1.44$ or $ 1.57<\abs{\eta_\mathrm{SC}}< 2.5$, where $\eta_\mathrm{SC}$ is determined by the cluster position in the ECAL with respect to the center of the CMS detector. This excludes the transition region between EB and EE. The \pt of each lepton has to be greater than 20\GeV, and the dilepton mass $M_{\ell\ell}$ is required to be greater than 50\GeV. At least one photon candidate with $\ptg > 15$\GeV is required. The $\eta$ range for photons is determined by the coverage of the ECAL and the silicon tracker and is the same as for electrons. The minimum distance between the photon and the leptons must be $\Delta R(\ell, \gamma) > 0.7$. In the very rare cases when a second photon is reconstructed, the photon with the higher \ptg is used for the differential cross section measurement.

\begin{figure}[tbh]
\centering
\includegraphics[width=0.49\textwidth]{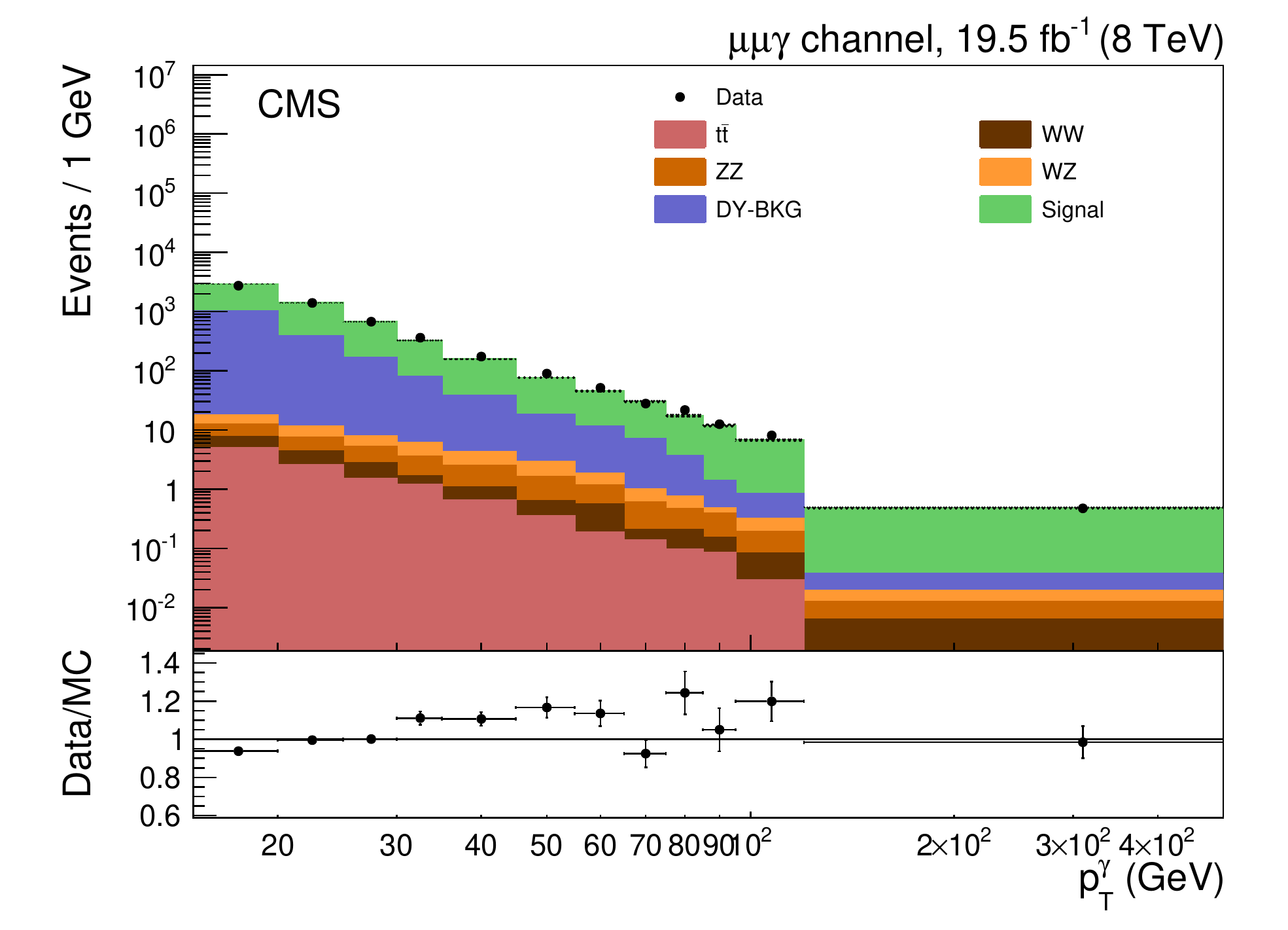}
\includegraphics[width=0.49\textwidth]{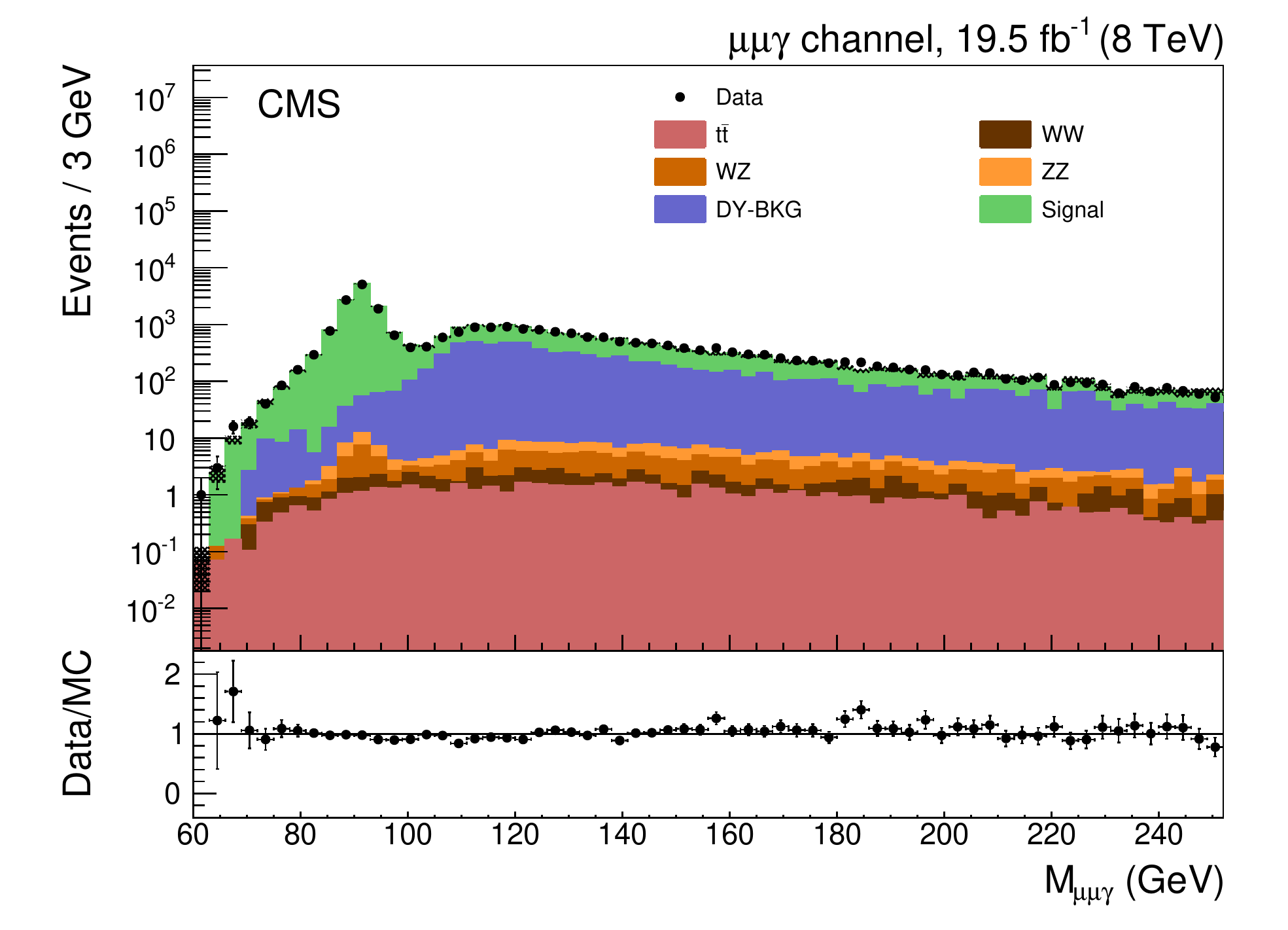}\\
\includegraphics[width=0.49\textwidth]{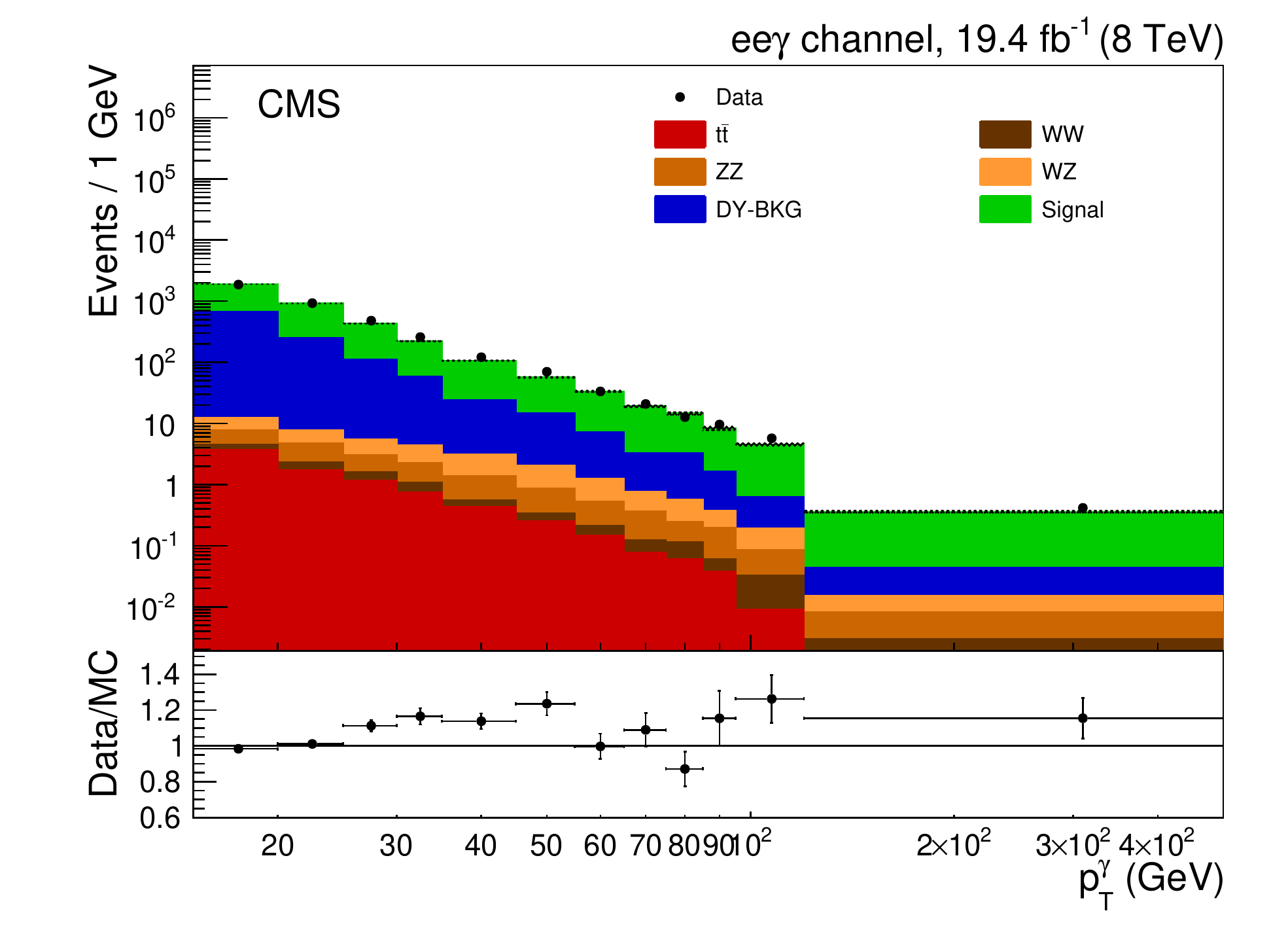}
\includegraphics[width=0.49\textwidth]{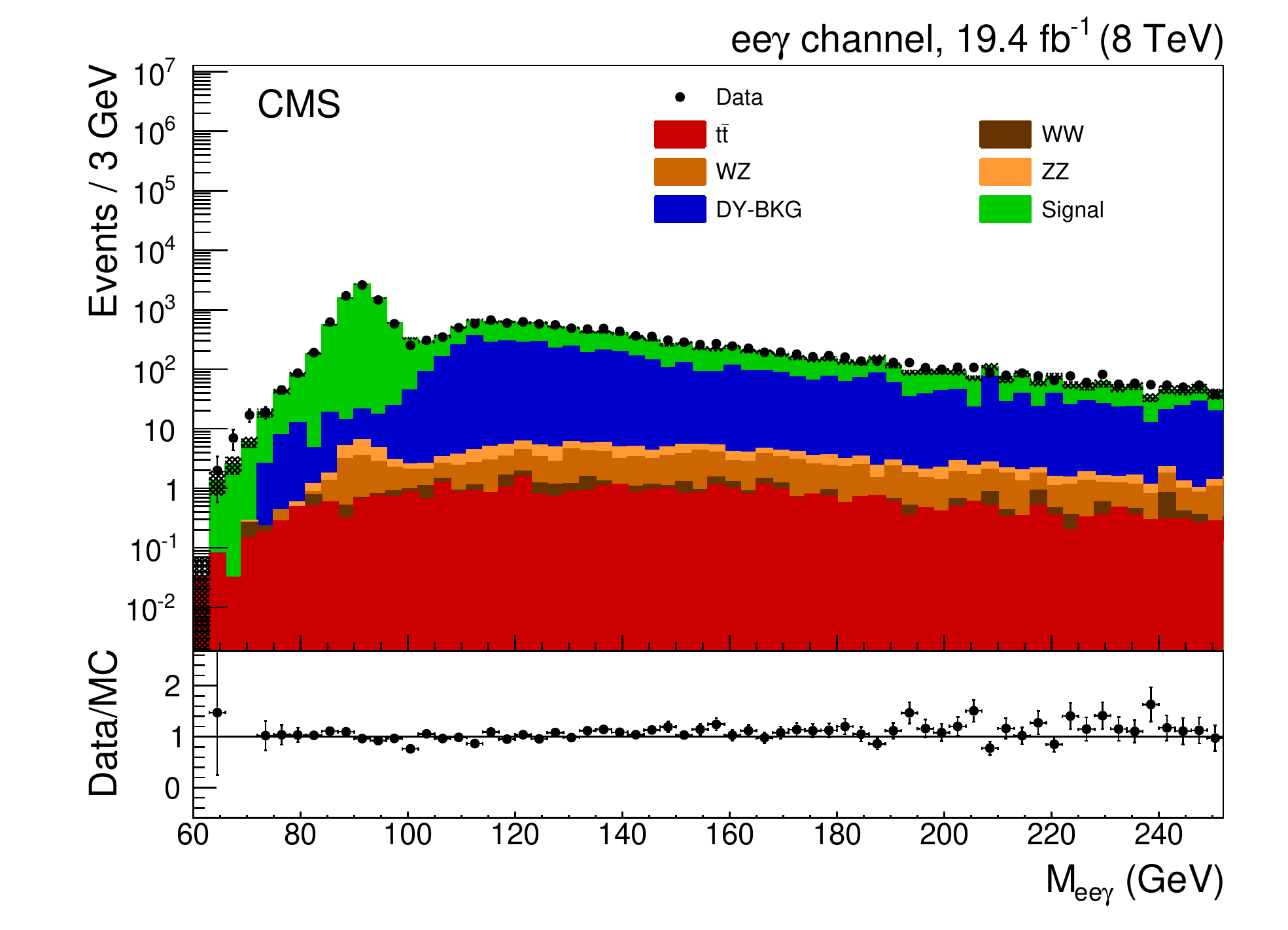}
\caption{Left: $\ptg$ distribution after the full event selection compared to the \SM prediction. Right: the distribution of the invariant mass $M_{\ell\ell\gamma}$. The mass distribution has a peak at the \Z boson mass, which arises from \FSR photons. Events with ISR photons appear in the large tail above the \Z boson mass where a large fraction of background is expected. The displayed uncertainties include only the statistical uncertainties in data and simulated samples.}
\label{PSEL1}
\end{figure}

A tag-and-probe method, similar to that presented in Ref.~\cite{CMS:2011aa}, is used to measure the lepton reconstruction efficiencies. The photon reconstruction efficiency is determined with a modified tag-and-probe method that makes use of the \Z boson mass peak in the $M_{\mu\mu\gamma}$ distribution for \FSR photons. Scale factors are obtained from the measured efficiencies to correct the simulation. In \FIG{PSEL1} the observed $\ptg$ distribution and the invariant mass of the two leptons and the photon candidate $M_{\ell\ell\gamma}$ are compared to the \SM expectation. The level of agreement is discussed in Section~\ref{TEMPMETHOD}.

\section{Background estimation}
\label{TEMPMETHOD}

The dominant background for this measurement is DY+jets containing nonprompt photons, e.g., through $\pi^0$ or $\eta$ decays, or hadrons misidentified as photons. A template method is used to estimate this background from data. This method relies on the power of an observable to discriminate between signal photons and background. The signal yield is obtained from a maximum-likelihood fit to the observed distribution of such an observable using the known distributions (``templates'') for signal photons and background.

The cross sections are measured with two different template observables. One template method uses the shower shape variable $\see$. The separation between the two photons from the decay of light particles such as $\pi^0$, albeit small, leads to a larger $\see$ value than for single photons. The shower width of strongly interacting particles that mimic a photon signature also tends to be larger.

The other template observable is $\nfp$, which is the \pt sum of all PF photon objects in a cone of $\Delta R < 0.4$ around the photon. The abbreviation ``nfp" stands for ``no footprint of the photon" and indicates the removal of energy associated with the photon from the isolation variable. This energy is removed by excluding all particles whose ECAL clusters overlap with the photon cluster from the isolation variable. This makes $\nfp$, and hence the signal template, independent of \ptg. Since the background particles are often produced in cascade decays, $\nfp$ for them is expected to be greater on average than for signal photons.

\subsection{Extraction of signal and background templates}
The signal templates for both template variables are taken from data. A sample of photons with a background contamination of less than 1\% is obtained from {\FSR} \Zg events. About 23 thousand photon candidates close to one of the leptons, with $0.3 < \Delta R(\ell, \gamma) < 0.8$, are selected. The minimum separation is chosen to reduce the influence of the lepton on the template variables. The invariant dilepton mass is selected to be between 40 and 80\GeV. One lepton is required to have a $\pt > 30$\GeV while for the other lepton only $\pt > 10$\GeV is required. All photon quality requirements are applied except for $\see$, which cannot be used in the photon selection for the $\see$ template method. For the $\nfp$ template method the selection on $\Ig$ is not applied since the two variables are strongly correlated.

Different templates are chosen for the EB and EE regions, as well as for the lower $\ptg$ bins of the cross section measurement. Due to the limited number of photons a common template is used for $\ptg > 35$\GeV. The uncertainties in the signal templates are discussed in Section~\ref{SE}.

For the background templates it is almost impossible to find a sample of nonprompt or misidentified photons that is free of signal-like prompt photons. Therefore, we choose a jet data sample where such background objects are enhanced. From this sample events with two leading hadronic jets with $\pt > 30$\GeV and no isolated muon or electron are selected. Additionally, we require a photon candidate with a minimum separation from the jets of $\Delta R (\gamma, \mathrm{jet}) > 0.7$.  Kinematic distributions of the jets and the photon candidates as well as the distributions of the photon selection variables in the jet data sample are well described by the QCD simulation. This agreement allows us to establish a selection of photon candidates for the background template using the QCD simulation and to apply the same selection on the jet data sample. This selection is required to be dominated by nonprompt and misidentified photon candidates whose template shape is in agreement with the background template prediction from the DY+jets simulation. When defining the selection for the $\see$ background template, the photon candidates in the QCD simulation have to pass the full selection except for the requirements on $\see$ and $\Ic$. Starting from this preselection, the lower and upper boundaries on $\Ic$ are varied until a selection is found for which the template shape agrees with that in the DY+jets simulation. Once the selection is defined, it is applied to the jet data sample to obtain the $\see$ background template that is used for the signal extraction.

The same method is used to find an $\nfp$ background template. In this case, the photon preselection in the QCD simulation does not include the requirements on $\see$ and $\Ig$. Here the lower and upper boundaries on $\see$ are varied to find an appropriate selection for a $\nfp$ background template.

We use these methods to obtain background templates from the data for the various $\ptg$ bins in the EB and EE regions. The two different, almost uncorrelated, template variables are used for the cross section measurement and their results are compared. The methods rely on the DY+jets simulation, which is used to find the optimal background template selections. Hence, the agreement of the two methods provides an important consistency check. The uncertainties in these methods are discussed in Section~\ref{SE}.

\subsection{Signal extraction}
\label{SE}
\begin{figure}[tbp]
\centering
\includegraphics[width=0.49\textwidth]{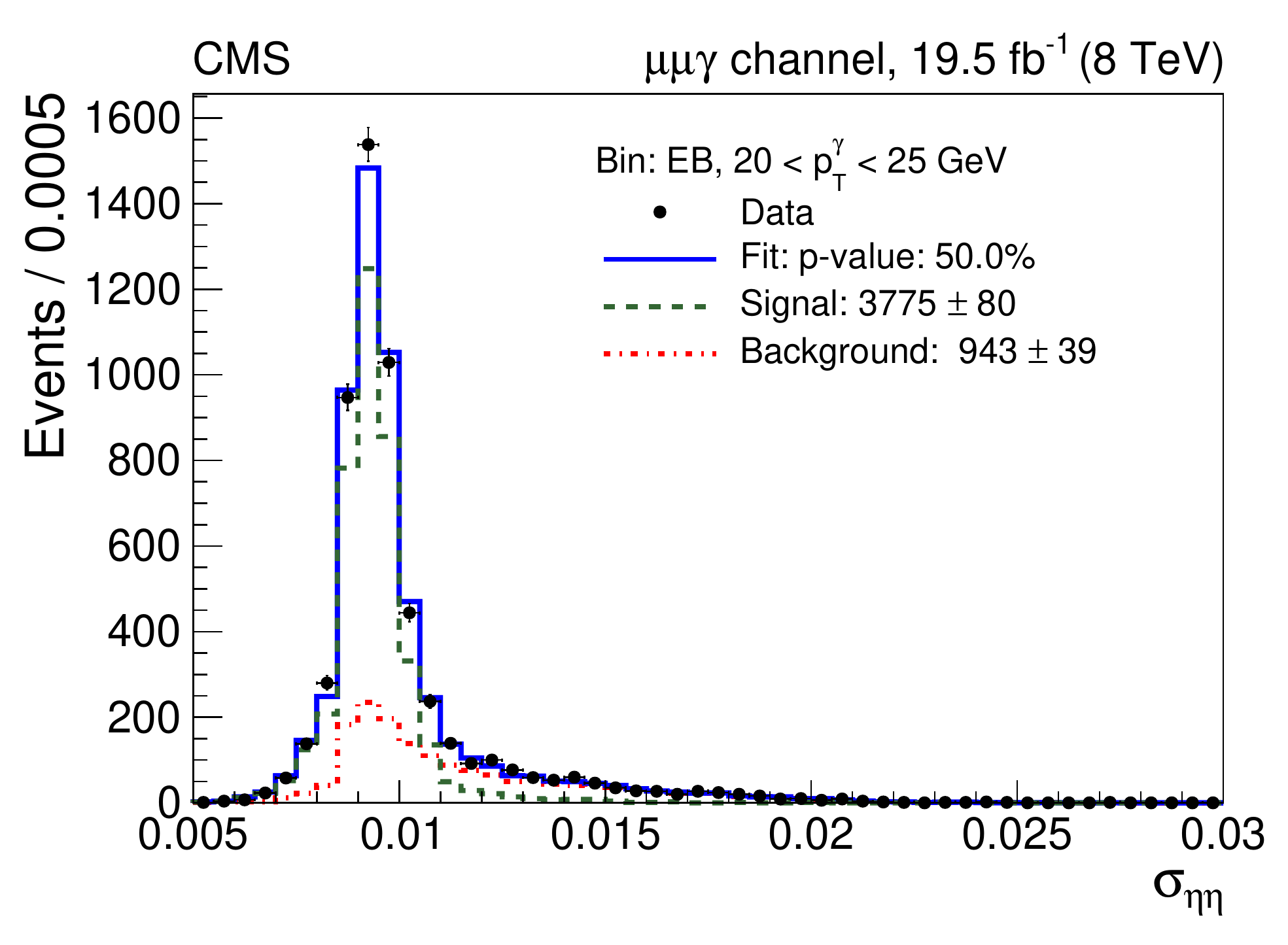}
\includegraphics[width=0.49\textwidth]{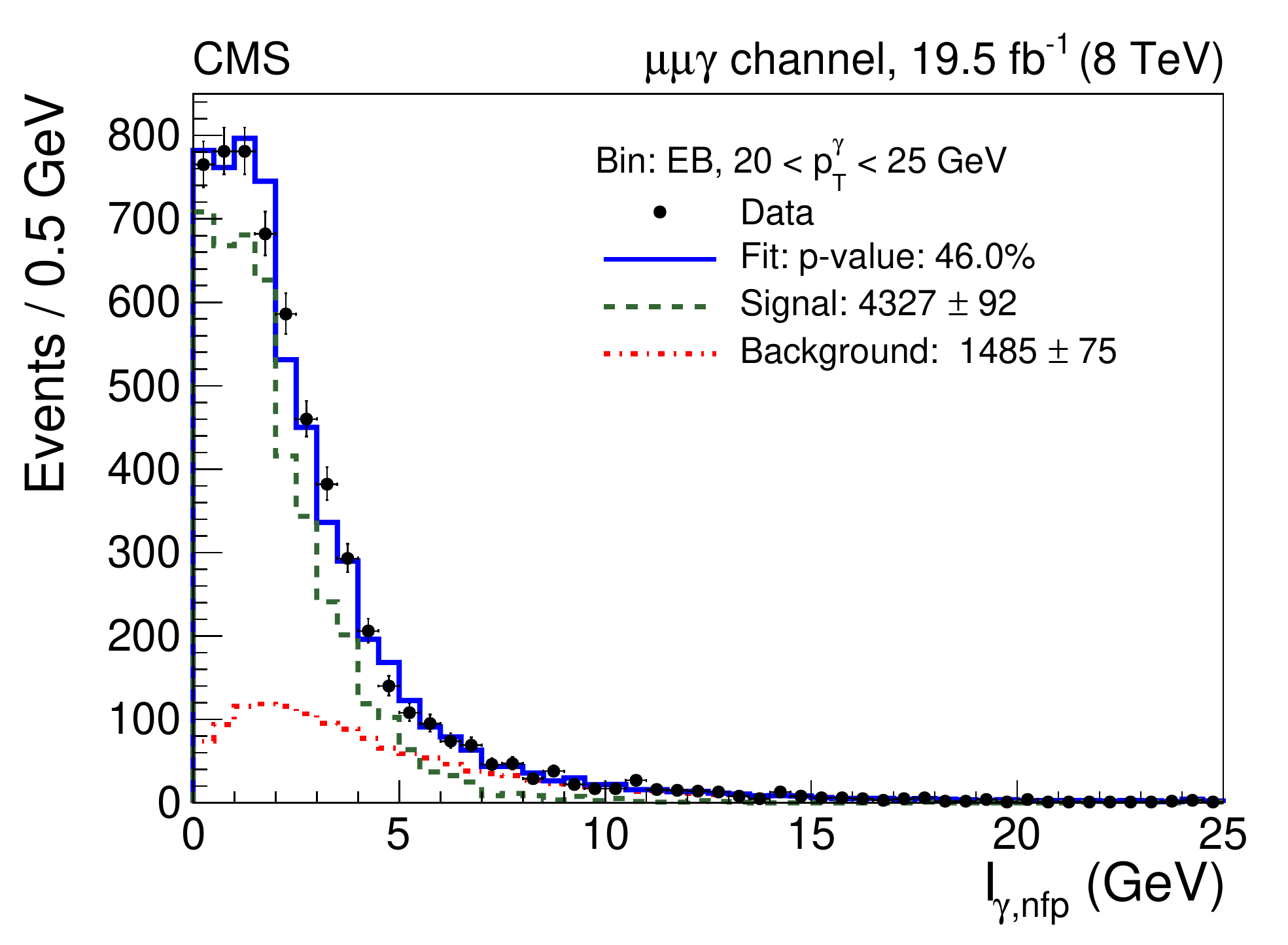}\\
\caption{Fits of the $\see$ templates (left) and the $\nfp$ templates (right) to the data for $20  < \ptg < 25$\GeV in the EB region. The extracted signal contributions are indicated by the green curves and the background contributions by the red ones.}
\label{PSE1}
\end{figure}

Using the templates obtained from the procedure described above the number of signal events is extracted in twelve $\ptg$ bins, separately for the EB and EE. Examples of these binned maximum-likelihood fits are shown in~\FIG{PSE1}. For the $\see$ template method the $\see$ requirement of the photon selection is applied after the fit. For the $\nfp$ template method the selection on $\Ig$ cannot be applied on the binned data after the fit. Consequently, the photon selection efficiencies are different for the two methods. Therefore, we should not expect the same number of signal and background events before corrections for the efficiencies are applied.

The following sources of uncertainties are considered for the signal extraction:
\begin{itemize}
\item The statistical uncertainty in the signal templates due to the limited number of \FSR photons available in data results in an uncertainty of 0.5--2\% (EB) and 0.5--9\% (EE) in the extracted signal yield that increases from low to high $\ptg$.
\item  The systematic uncertainty in the signal templates due to contamination of the \FSR sample and the usage of a common template for all bins with $\ptg > 35$\GeV is estimated using the simulation. The uncertainties are 0.5--3\% (EB) and 0.5--12\% (EE) in the signal yield increasing from low to high $\ptg$.
\item The number of events available in the DY+jets simulation that are used to find the appropriate selection for the background templates is low, especially in bins at high $\ptg$. The uncertainty in the extracted signal yield due to this limited sample size is 0.6--3\% (EB) and 1.6--5\% (EE) increasing from low to high $\ptg$.
\item The agreement of the QCD simulation and jet data sample is essential for the background template determination. We evaluate the uncertainty due to this imperfect modeling by calculating the standard deviation of the difference between the signal fraction obtained with template fits in data and simulation for a large number of different background template selections each defined by certain lower and upper boundaries on the template selection variable, $\Ic$ for the $\see$ templates and $\see$ for the $\nfp$ templates. For data the background templates are taken from the jet data sample and for simulation from the QCD simulation sample. The uncertainty is estimated to be 0.3--6\% (EB) and 3--6\% (EE) increasing from low to high $\ptg$.
\item The statistical uncertainty in the signal yield obtained from the template fit is very similar for the two methods and amounts to 2--9\% (EB) and 3--14\% (EE) increasing from low to high $\ptg$.
\end{itemize}
Additionally, we have to consider the small fraction of irreducible background events from $\ttbar$, ZZ, ZW, and WW production. These background yields are estimated from the \SM simulation and subtracted from the \ptg distribution of signal candidates. At low $\ptg$ this contribution is negligible compared to the background from nonprompt or misidentified photons. At higher $\ptg$ the fraction of irreducible background events is less than 4\%, which is small compared to the overall uncertainty of the measurement. Since these backgrounds are very small their uncertainties have a negligible effect on the measurement.

\section{Cross section measurement}
\label{UNF}
We measure the cross section for a phase space region that corresponds closely to that used for the event selection. This phase space is defined by several kinematic requirements on the final-state particles: the leptons from the \Z boson decay need to satisfy $\pt > 20$\GeV and $\abs{\eta} < 2.5$, and the dilepton mass has to be greater than 50\GeV. The photon is required to have $\abs{\eta} < 2.5$ and needs to be separated from both leptons by $\Delta R(\ell, \gamma) > 0.7$. Finally, a requirement is put on the photon isolation at the generator level $I_\text{gen} < 5$\GeV to exclude photons from jet fragmentation. The isolation variable uses a cone size of $\Delta R < 0.3$ and sums the transverse momentum of partons in the case of \MCFM and of final-state particles in the case of \SHERPA. It has been verified with \SHERPA that photons that do not pass the $I_\text{gen}$ selection also fail to pass the photon selection at the detector level. The definition of the phase space for the cross section measurements is summarized in \TAB{TXS1}.

The procedure for extracting the differential cross section from the number of observed signal events involves two steps. First, we extract the number of events produced in each \ptg bin within a phase space defined by the requirements in~\TAB{TXS1} and the additional experimental requirements on $\eta$ and $\eta_{\mathrm{SC}}$ as described in Section~\ref{SEL}. The number of observed signal events needs to be corrected for detector effects. These include efficiencies as well as bin migrations due to resolution and energy calibration. Both effects are treated using unfolding techniques. For the unfolding the method of D'Agostini~\cite{D'Agostini:1994zf}, as implemented in the \textsc{RooUnfold}~\cite{RooUnfold} software package, is used. The response matrices are obtained from the signal simulation. A different response matrix is required for the two template methods because of the different photon selections. After the unfolding, compatible signal yields are obtained with the two template methods. Bias and variance of the unfolding procedure are estimated using pseudoexperiments. The uncertainties in the unfolding are 1\% at low $\ptg$ increasing up to 6\% for the high-$\ptg$ bins. To estimate the effect of the uncertainties in the photon energy scale and resolution, the unfolding of the data is repeated varying the photon energy scale and resolution in the response matrix within one standard deviation. The observed effect on the unfolded event yield is about 2.4\% and is almost independent of $\ptg$.

The second step is to extrapolate the unfolded event yield in each \ptg bin to the desired phase space taking into account the detector acceptance, which is calculated using \MCFM (NLO) and verified with \SHERPA. About 92\% of the muon channel events and 87\% of the electron channel events are within the detector acceptance. These values are only slightly \ptg dependent.

{\renewcommand{\arraystretch}{1.2}
\begin{table}[bh]
\caption{Phase space definition of the \Zg cross section measurements.}
\begin{center}
\begin{tabular}{l}
\hline
Cross section phase space\\\hline
$M_{\ell\ell} > 50$\GeV\\
$\Delta R(\ell, \gamma) > 0.7$\\
photon: $\abs{\eta} < 2.5$, $I_\text{gen} < 5$\GeV\\
leptons: $\abs{\eta} < 2.5$, $\pt > 20$\GeV\\
\label{TXS1}
\end{tabular}
\end{center}
\end{table}
}

\subsection{The inclusive cross section}
\label{XS}
\begin{figure}[tbp]
\centering
\includegraphics[width=0.49\textwidth]{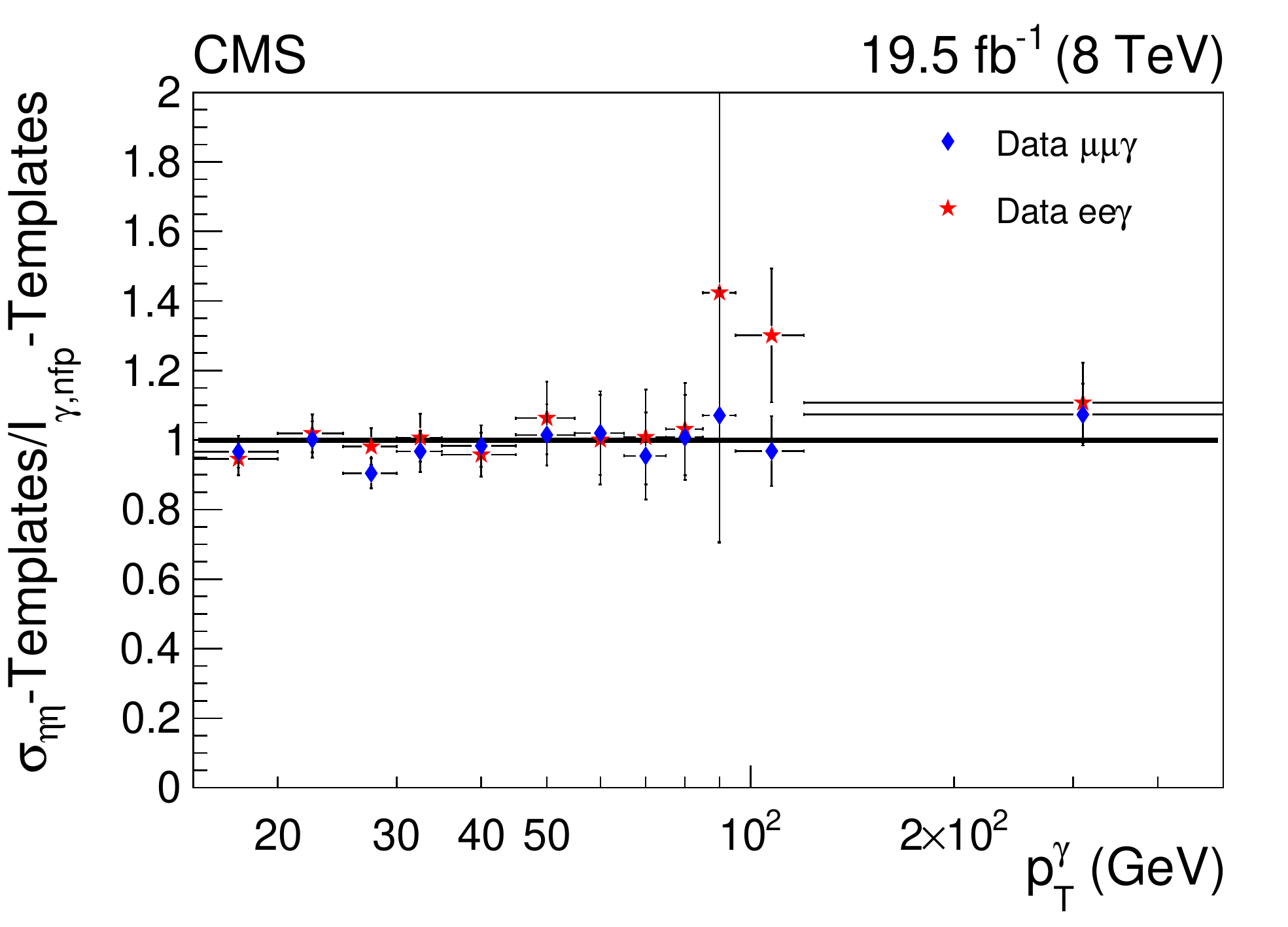}
\includegraphics[width=0.49\textwidth]{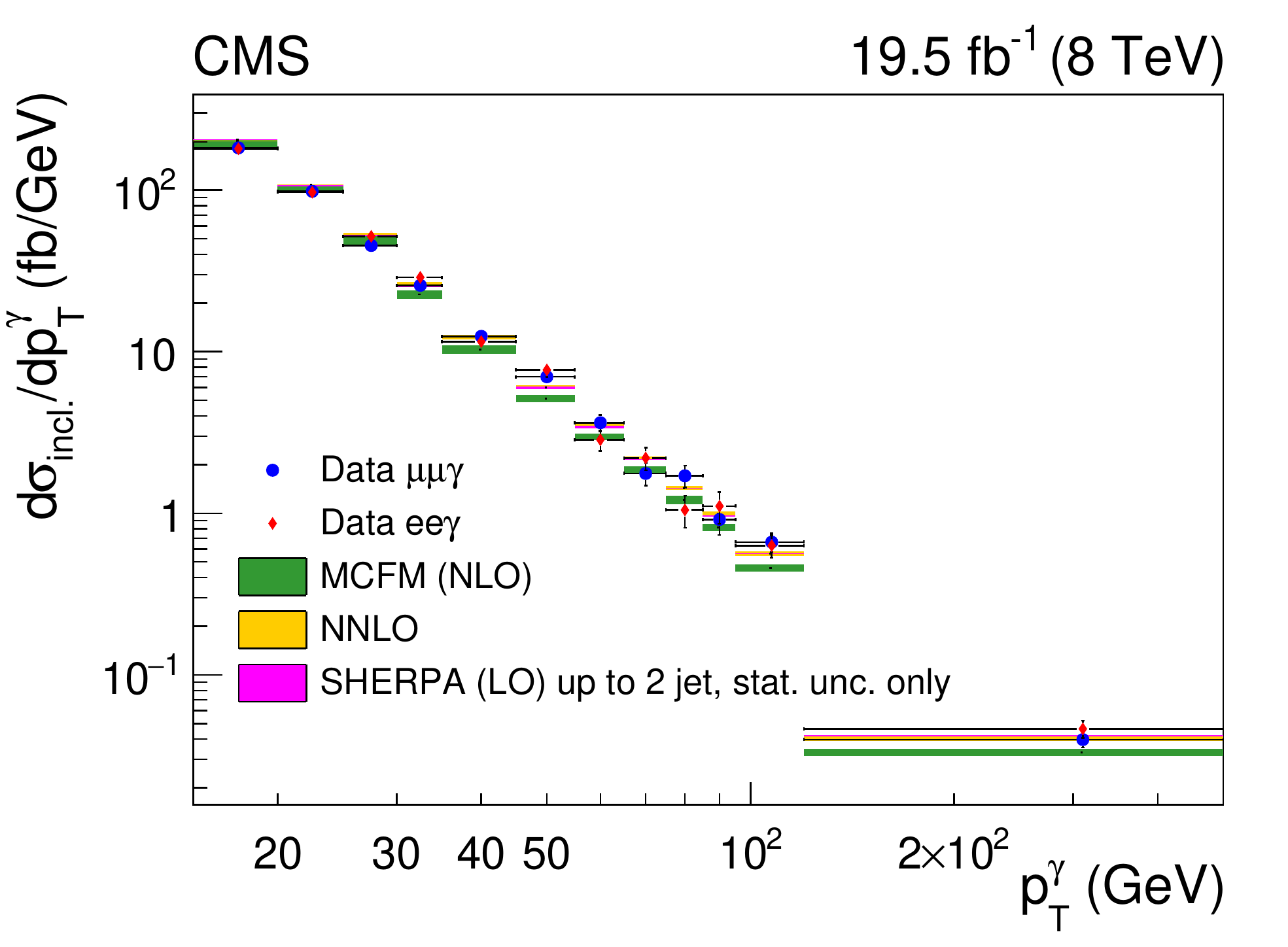}
\caption{Left: ratio of the inclusive cross sections as obtained with the two template methods. Right: measured differential cross sections for the muon and electron channels using a combination of the two template methods compared to the NNLO~\cite{Grazzini:arXiv1309.7000}, the \MCFM (NLO) and the \SHERPA \SM predictions. The last bin is computed for the interval 120--500\GeV.}
\label{PXS2}
\end{figure}

The cross sections are calculated from the unfolded number of signal events $N_i$ and the detector acceptance $A_i$ in each $\ptg$ bin using the relation $\sigma_i = N_{i}/(A_i L)$ with an integrated luminosity of $L = 19.5\pm0.5$\fbinv for the muon channel and $L = 19.4\pm0.5$\fbinv for the electron channel.
The cross section values obtained with the two template methods are compatible within their uncertainties as shown in~\FIG{PXS2}~(left). The correlation between the template variables $\see$ and $\nfp$ is less than 30\%. The compatibility of the two results is a good indication that the background estimation is correct. The correlation of 30\% is also assumed for the uncertainties in the background subtractions with the two template methods. All other uncertainties, \ie, in the dilepton (2\%) and photon (2\%) efficiencies, the photon energy scale and resolution (2.4\%), unfolding (1--6\%), luminosity (2.6\%), and statistical uncertainties are assumed to be 100\% correlated between the two template methods. Since the two template methods show good agreement, the results are combined using the best linear unbiased estimator (BLUE) method~\cite{Valassi2003391}, which takes into account the correlation of all uncertainties.

\begin{figure}[tbp]
\centering
\includegraphics[width=0.7\textwidth]{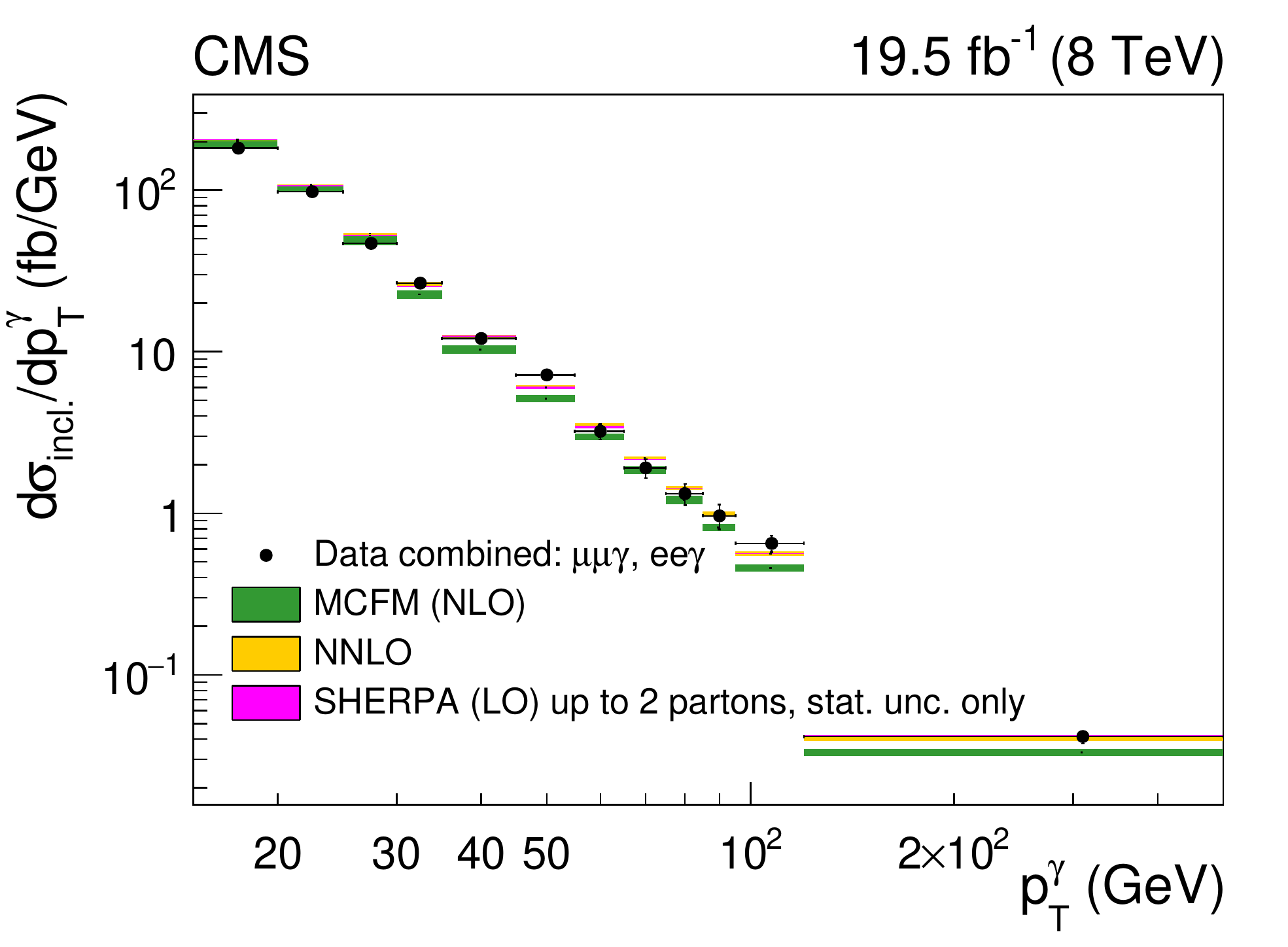}
\includegraphics[width=0.7\textwidth]{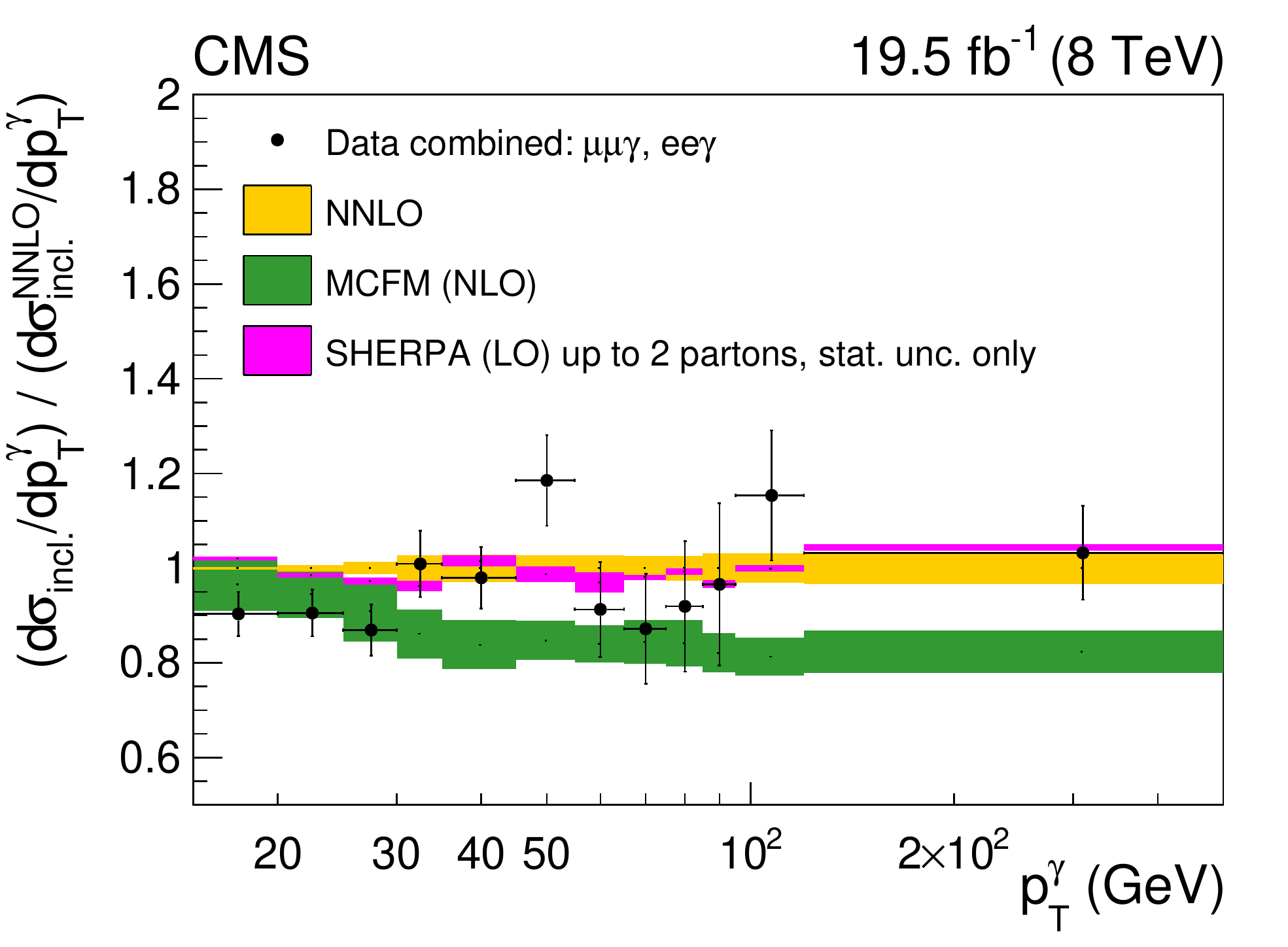}
\caption{Top: combined inclusive differential cross section for the two lepton channels compared to the NNLO~\cite{Grazzini:arXiv1309.7000}, the \MCFM (NLO) and the \SHERPA \SM predictions. The latter is normalized to the NNLO cross section. The last bin is computed for the interval 120--500\GeV. Bottom: the ratios of the data and the other predictions to the NNLO calculation showing the effect of the additional partons on the inclusive cross section.}
\label{PXS3b}
\end{figure}

The combined results of the two template methods for the muon and the electron channels are, as expected from lepton universality, fully compatible as presented in~\FIG{PXS2}~(right). For $\ptg > 15$\GeV inclusive cross sections of $2066 \pm 23\stat\pm97\syst\pm 54\lum\fb$ and $2087 \pm 30\stat\pm 104\syst\pm54\lum\fb$ are measured for the muon and electron channels, respectively. The cross sections are combined using the BLUE method~\cite{Valassi2003391}, assuming that the systematic uncertainties between the two lepton channels are highly correlated, since the signal yields are extracted using the same template shapes. The combined cross sections for the two channels are given in~\TAB{TXS3} and shown as the differential cross section in~\FIG{PXS3b}. It is compared to the \MCFM (NLO), the NNLO, and the \SHERPA predictions. For $\ptg > 15$\GeV the inclusive cross section is measured to be \begin{center}
$\sigma_\text{incl}= 2063\pm 19\stat\pm 98\syst\pm 54\lum\fb$.
\end{center}
This is in good agreement with the \MCFM prediction of $\sigma_\text{incl}^{\MCFM} =  2100\pm 120\fb$ and the NNLO calculation~\cite{Grazzini:arXiv1309.7000} of $\sigma_\text{incl}^{\mathrm{NNLO}} =  2241\pm 22\,\text{(scale only)}\fb$. However, the ratio plot in \FIG{PXS3b} shows that at high \ptg the measurement is better described by the NNLO calculation and by \SHERPA than by \MCFM. The \SHERPA calculation includes up to two partons in the matrix element which leads to a significant enhancement at high \ptg.

{\renewcommand{\arraystretch}{1.2}
\begin{table}[tp]
\caption{The combined inclusive cross sections for the muon and electron channels with statistical, systematic, and integrated luminosity uncertainties, respectively. Scale and PDF uncertainties are included in the systematics for the \MCFM (NLO) cross section calculation. Only scale uncertainties are considered for the NNLO calculation.}
\begin{center}
\begin{tabular}{c|ccc}
\hline
\ptg (\GeVns{}) & $\sigma_\text{incl}$\,(fb) & $\sigma_\text{incl}^{\MCFM}$\,(fb) & $\sigma_\text{incl}^{\mathrm{NNLO}}$\,(fb) \\\hline
15--20 &$ 908\pm 12\pm 39\pm 24 $&$ 972\pm 57 $&$ 1005.6\pm 2.6$\\
20--25 &$ 489\pm 9\pm 21\pm 13 $&$ 510\pm 27 $&$ 540.1\pm 3.7$\\
25--30 &$ 234\pm 7\pm 11\pm 6 $&$ 245\pm 17 $&$ 269.2\pm 3.6$\\
30--35 &$ 132.8\pm 4.8\pm 7.0\pm 3.5 $&$ 113.4\pm 6.8 $&$ 131.6\pm 3.5$\\
35--45 &$ 120.7\pm 4.0\pm 6.2\pm 3.1 $&$ 103.2\pm 6.4 $&$ 123.2\pm 3.6$\\
45--55 &$ 71.8\pm 3.0\pm 4.6\pm 1.9 $&$ 51.3\pm 2.5 $&$ 60.6\pm 1.6$\\
55--65 &$ 32.2\pm 2.3\pm 2.5\pm 0.8 $&$ 29.6\pm 1.4 $&$ 35.2\pm 1.0$\\
65--75 &$ 19.1\pm 1.8\pm 1.7\pm 0.5 $&$ 18.5\pm 1.0 $&$ 21.89\pm 0.56$\\
75--85 &$ 13.2\pm 1.5\pm 1.2\pm 0.3 $&$ 12.10\pm 0.70 $&$ 14.38\pm 0.38$\\
85--95 &$ 9.6\pm 1.2\pm 1.2\pm 0.3 $&$ 8.19\pm 0.41 $&$ 9.98\pm 0.31$\\
95--120&$ 16.3\pm 1.3\pm 1.4\pm 0.4 $&$ 11.47\pm 0.57 $&$ 14.10\pm 0.44$\\
$>$120 &$ 15.8\pm 1.0\pm 1.0\pm 0.4 $&$ 12.59\pm 0.68 $&$ 15.29\pm 0.51$
\label{TXS3}
\end{tabular}
\end{center}
\end{table}
}

\subsection{The exclusive cross section}
\label{XSE}

To understand the effect of additional jets a measurement of the exclusive cross section is performed for \Zg production without any accompanying jet with $\pt > 30$\GeV and $\abs{\eta} < 2.4$.

The high instantaneous luminosity in the 2012 run requires that special care must be taken to reduce the contribution from jets produced in pileup interactions. About 50\% of these jets can be rejected by requiring a maximal \pt fraction of charged particles in a jet originating from a pileup vertex. Further corrections are needed to account for the remaining contribution from pileup jets and jet reconstruction inefficiencies. The jet reconstruction efficiencies and jet misidentification rates for each \ptg bin are taken from the simulation where the jet misidentification rate considers all jets that cannot be matched to a jet from the main interaction at generator level. These are used to calculate the number of exclusive events from the measured number of inclusive events and the measured number of events with zero reconstructed jets. The latter are determined with the same methods used for the extraction of the inclusive signal yield. The uncertainties in the cross section due to pileup and jet energy scale are evaluated to be 1\% and 2.5\% respectively.

\begin{figure}[tbp]
\centering
\includegraphics[width=0.49\textwidth]{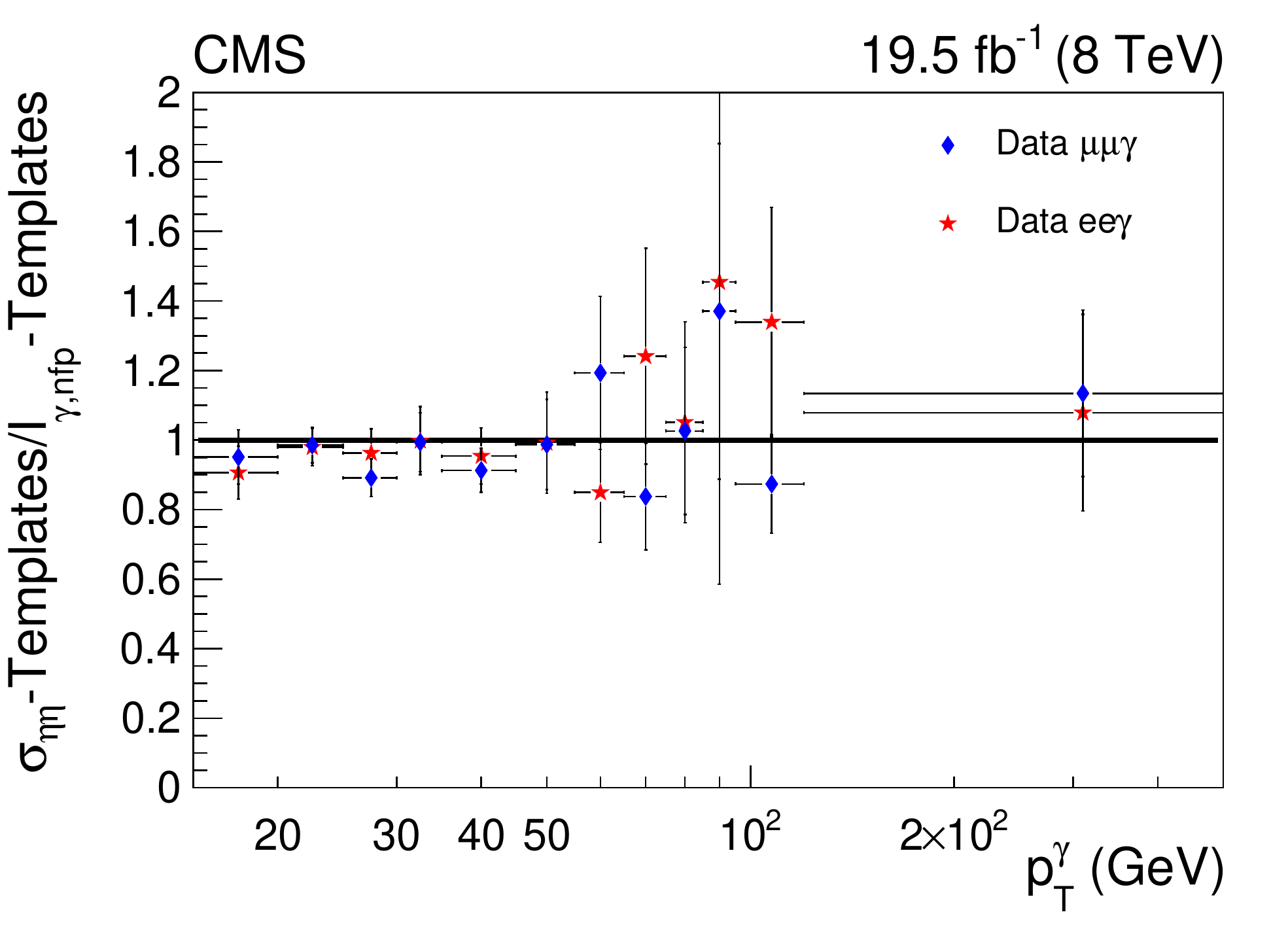}
\includegraphics[width=0.49\textwidth]{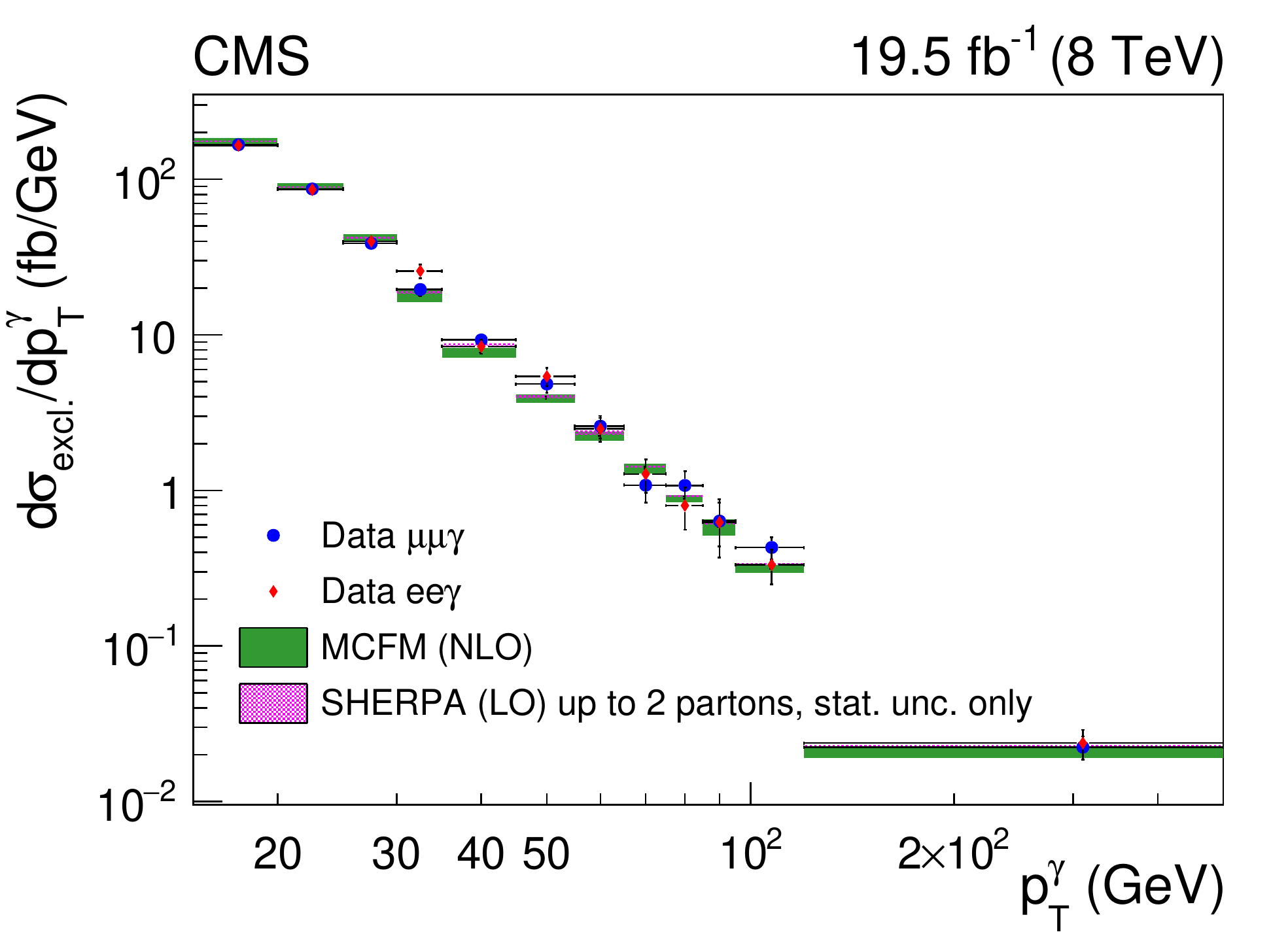}
\caption{Left: ratio of the exclusive cross sections as obtained with the two template methods. Right: measured cross sections for the muon and electron channels using a combination of the two template methods compared to the \MCFM (NLO) and the \SHERPA \SM predictions. The last bin is computed for the interval 120--500\GeV.}
\label{PXSE3}
\end{figure}

\begin{figure}[tbp]
\centering
\includegraphics[width=0.7\textwidth]{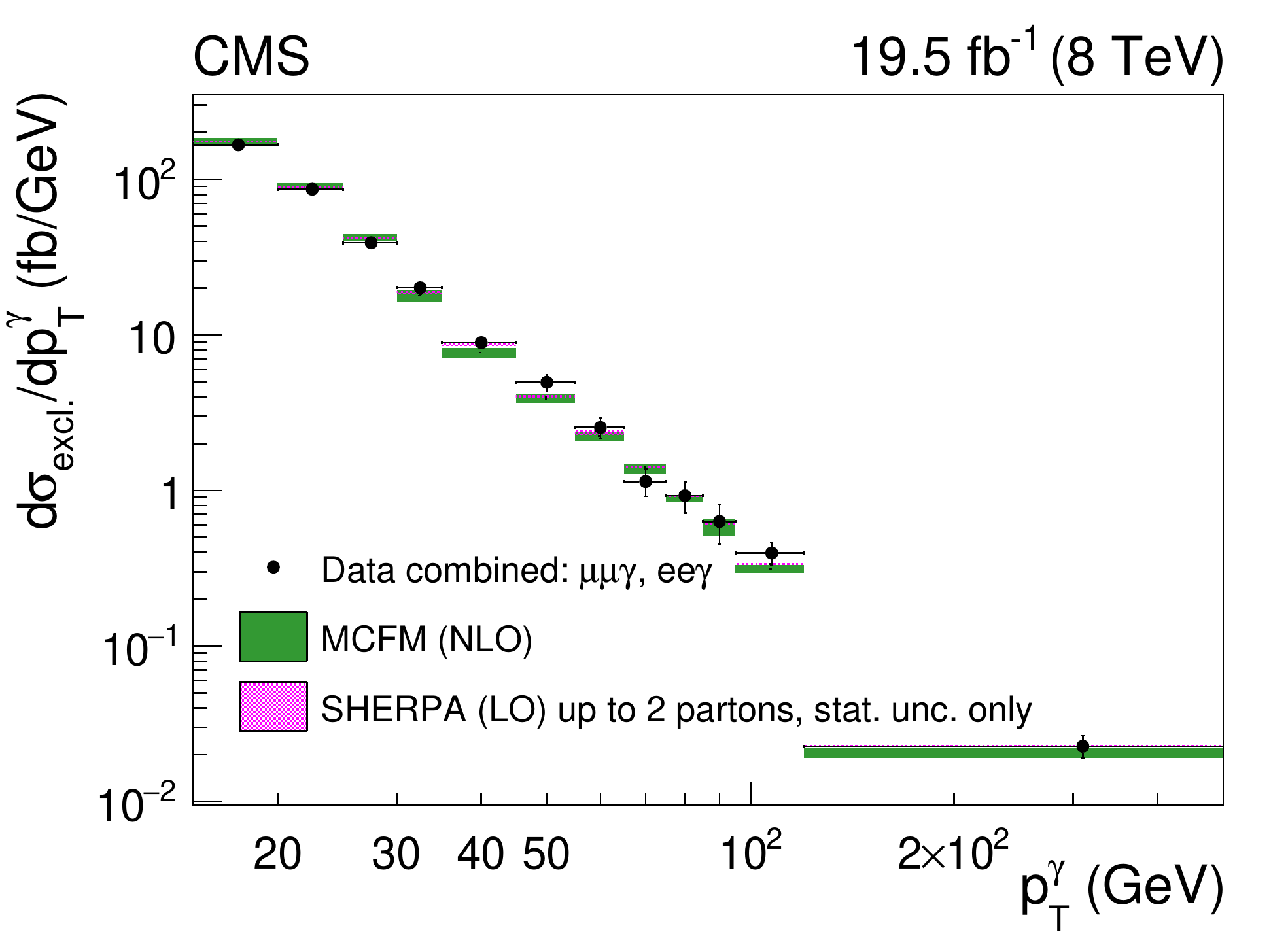}
\includegraphics[width=0.7\textwidth]{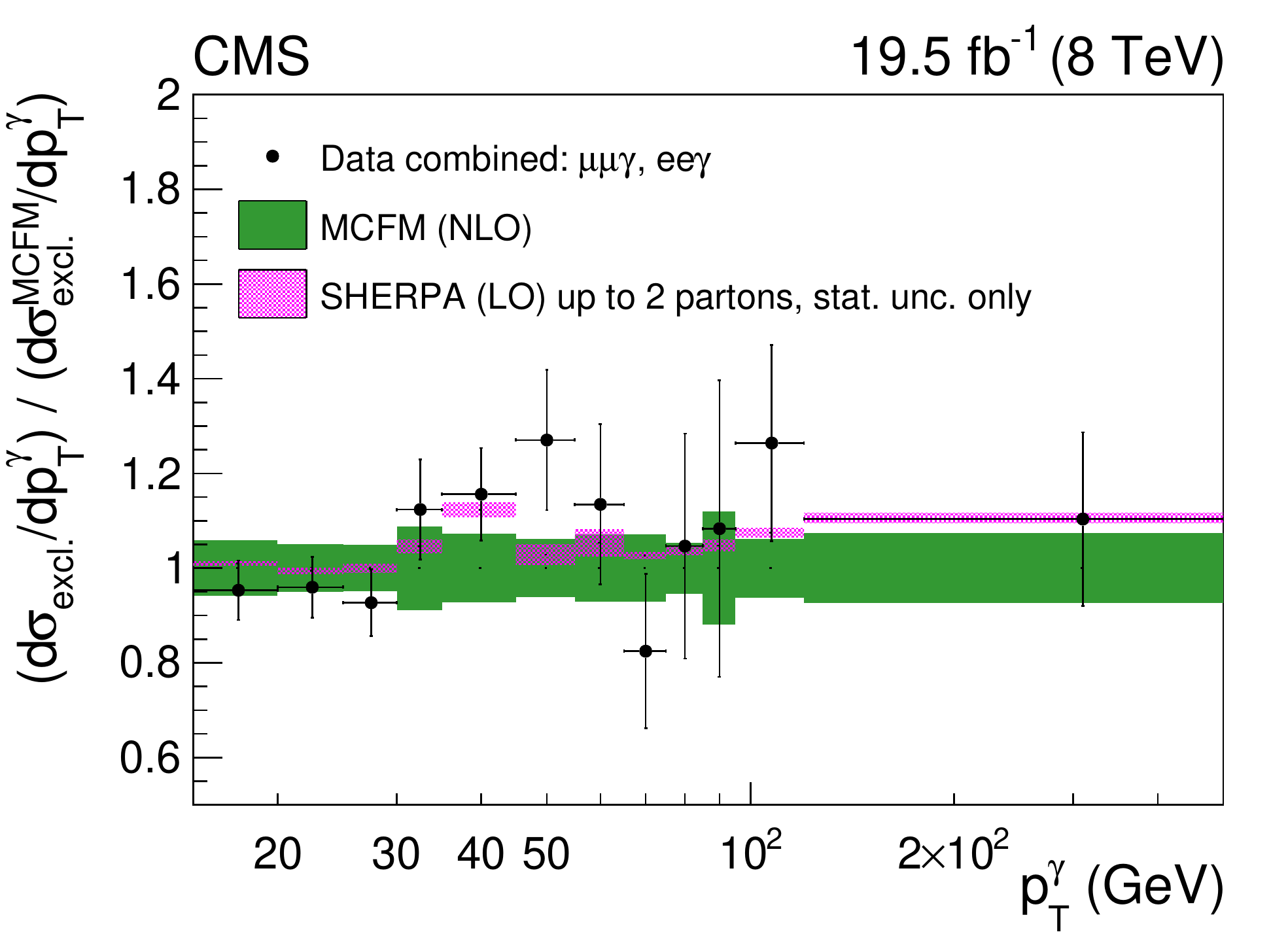}
\caption{Top: combined exclusive differential cross section for the two lepton channels compared to the \MCFM (NLO) and \SHERPA \SM predictions. The whole \SHERPA sample (inclusive) is normalized to the NNLO cross section. The last bin is computed for the interval 120--500\GeV. Bottom: for the exclusive measurement the ratios to the \MCFM (NLO) prediction are shown.}
\label{PXSE3b}
\end{figure}

The $\ptg$ distribution of exclusive events is unfolded and the cross sections are calculated. The acceptance is taken from \MCFM (NLO). As shown in~\FIG{PXSE3}~(left) the results of the two template methods agree well and are combined using the BLUE method. With the requirement of $\ptg > 15$\GeV for the muon and the electron channels exclusive cross sections of $1774\pm 23\stat\pm 115\syst\pm 46\lum\fb$ and $1791\pm 29\stat\pm 122\syst\pm 47\lum\fb$ are measured, respectively. These and the differential cross sections presented in~\FIG{PXSE3}~(right) are compatible. The combined cross sections for the two channels are shown in \FIG{PXSE3b}.

\begin{figure}[tbp]
\centering
\includegraphics[width=0.7\textwidth]{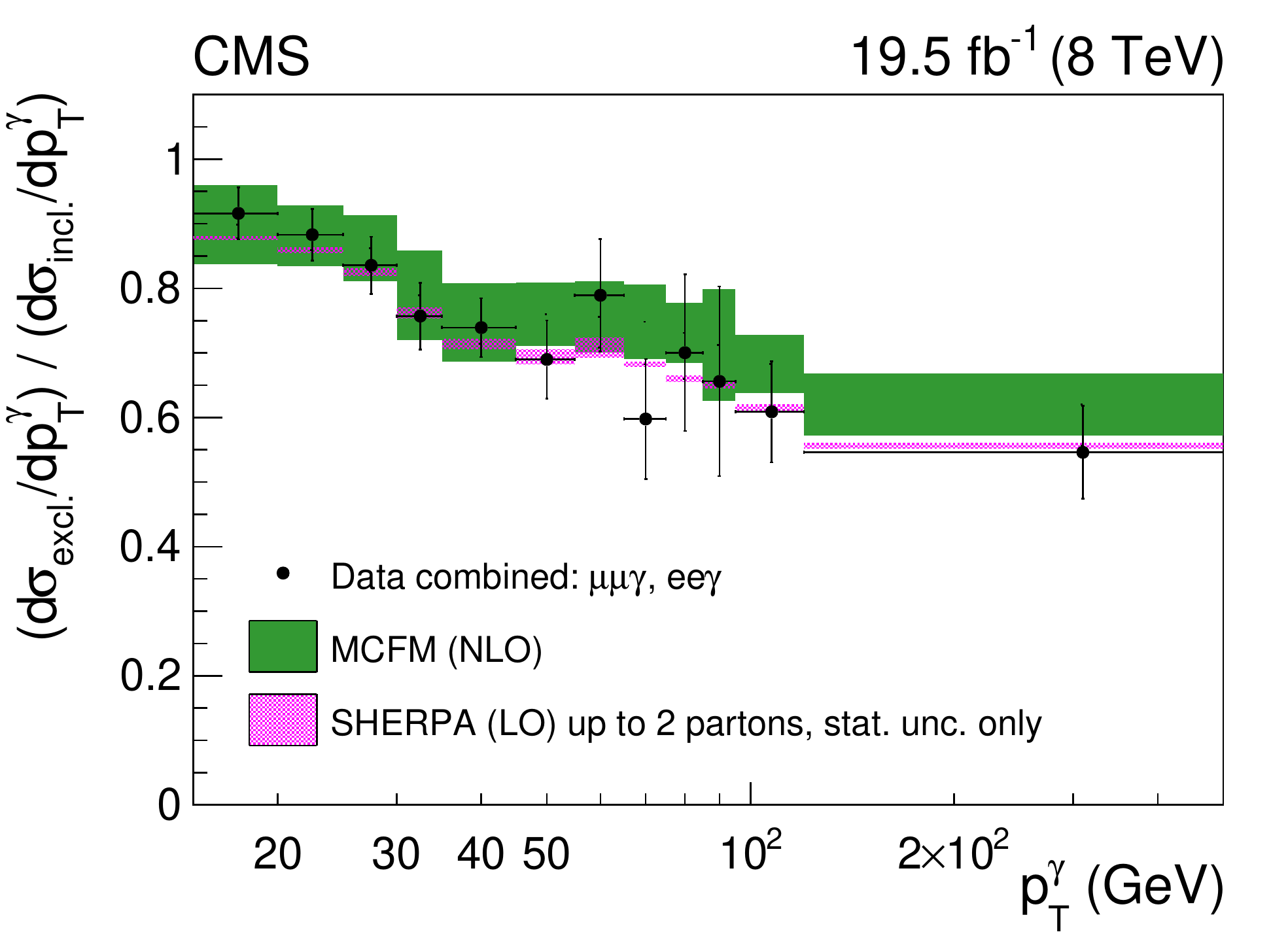}
\caption{Ratio of the exclusive to inclusive cross sections for \Zg production.}
\label{PXSE4}
\end{figure}

The difference at high $\ptg$ between the \MCFM (NLO) calculation and \SHERPA with up to two partons is smaller for the exclusive calculation. The measured cross section values are in agreement with the two predictions. The combination of the two lepton channels compared to \MCFM (NLO) is presented in \TAB{TXSE5} and the differential cross section is shown in \FIG{PXSE3b}. The ratio of the exclusive and inclusive cross sections is shown in \FIG{PXSE4}. The fraction of exclusive events decreases with increasing $\ptg$ and the fraction of events with additional jets changes from 10\% to 40\%. Adding the exclusive cross sections in all bins we obtain for $\ptg > 15$\GeV
\begin{center}
$\sigma_{\text{excl}}= 1770\pm 18\stat\pm 115\syst\pm 46\lum\fb$.
\end{center}
This is compatible with the \MCFM (NLO) prediction of $\sigma_{\text{excl}}^{\MCFM}= 1800\pm 120\fb$.

{\renewcommand{\arraystretch}{1.2}
\begin{table}[tp]
\caption{The combined exclusive cross sections for muon and electron channels with statistical, systematic, and luminosity uncertainties respectively. Scale and PDF uncertainties are included in the systematics for the \MCFM (NLO) cross section calculation.}
\begin{center}
\begin{tabular}{c|cc}
\hline
\ptg (\GeVns{}) & $\sigma_{\text{excl}}$\,(fb) & $\sigma_{\text{excl}}^{\MCFM}$\,(fb) \\\hline
15--20 &$ 832\pm 12\pm 49\pm 22 $&$ 873\pm 51$\\
20--25 &$ 432\pm 9\pm 25\pm 11 $&$ 450\pm 23$\\
25--30 &$ 196\pm 6\pm 12\pm 5 $&$ 211\pm 10$\\
30--35 &$ 100.5\pm 5.3\pm 7.4\pm 2.6 $&$ 89.5\pm 7.9$\\
35--45 &$ 89.2\pm 3.7\pm 6.2\pm 2.3 $&$ 77.2\pm 5.6$\\
45--55 &$ 49.5\pm 2.8\pm 4.9\pm 1.3 $&$ 39.0\pm 2.4$\\
55--65 &$ 25.4\pm 2.0\pm 3.1\pm 0.7 $&$ 22.4\pm 1.6$\\
65--75 &$ 11.4\pm 1.5\pm 1.7\pm 0.3 $&$ 13.83\pm 0.98$\\
75--85 &$ 9.3\pm 1.3\pm 1.6\pm 0.2 $&$ 8.85\pm 0.48$\\
85--95 &$ 6.3\pm 1.2\pm 1.4\pm 0.2 $&$ 5.83\pm 0.70$\\
95--120&$ 9.9\pm 1.0\pm 1.3\pm 0.3 $&$ 7.83\pm 0.48$\\
$>$120 &$ 8.6\pm 0.8\pm 1.1\pm 0.2 $&$ 7.81\pm 0.58$
\label{TXSE5}
\end{tabular}
\end{center}
\end{table}
}

\section{Limits on {\AGC}s}
\label{LIMITS}
The \ZZg or {\Zgg} {\AGC}s are formulated in the framework of an effective field theory considering dimension six and eight operators, that fulfill the requirements of Lorentz invariance and local U(1) gauge symmetry. The resulting Lagrangian \cite{PhysRevD.61.073013} has the form
\begin{equation}\begin{split}
\label{FAGC2}
    L_{\mathrm{aTGC}} = L_{\mathrm{SM}} + \frac{e}{m_\mathrm{Z}^2} & \left[ -[h_1^\gamma(\partial^\sigma F_{\sigma\mu}) + h_1^Z(\partial^\sigma Z_{\sigma\mu})]Z_\beta F^{\mu\beta} \right.\\
        & -[h_3^\gamma(\partial_\sigma F^{\sigma\rho}) + h_3^Z(\partial_\sigma Z^{\sigma\rho})]Z^\alpha \tilde{F}_{\rho\alpha}\\
        & -[\frac{h_2^\gamma}{m_\mathrm{Z}^2}[\partial_\alpha \partial_\beta \partial^\rho F_{\rho\mu}]  + \frac{h_2^Z}{m_\mathrm{Z}^2} [\partial_\alpha \partial_\beta (\partial_\nu\partial^\nu + m_\mathrm{Z}^2)Z_\mu]] Z^\alpha F^{\mu\beta}\\
        & \left. +[\frac{h_4^\gamma}{2m_\mathrm{Z}^2}[\partial_\nu\partial^\nu\partial^\sigma F^{\rho\alpha}] + \frac{h_4^Z}{2m_\mathrm{Z}^2}[(\partial_\nu\partial^\nu + m_\mathrm{Z}^2)\partial^\sigma Z^{\rho\alpha}]]Z_\sigma \tilde{F}_{\rho\alpha}\right]
\end{split}\end{equation}
with the electromagnetic tensor $F_{\mu\nu} = \partial_\mu F_\nu - \partial_\nu F_\mu$ and $\tilde{F}_{\mu\nu} = 1/2\,\epsilon_{\mu\nu\rho\sigma}F^{\rho\sigma}$ and similar definitions for the \Z boson field. There are eight coupling constants $h_i^V$, $i=1\dots4$ and $V = \Z,\gamma$ for \ZZg (\Zgg) couplings. The parameters $h_1^V$ and $h_2^V$ are CP-violating while $h_3^V$ and $h_4^V$ are not. It was shown in Ref.~\cite{Gounaris:hep-ph0005269, Degrande:arXiv1308.6323} that there is no dimension six operator respecting $U(1)_Y \times SU(2)_L$ invariance, but two dimension eight operators, including the Higgs field, that could lead to an enhancement proportional to $h_1^V$ and $h_3^V$. In this measurement we follow the CMS convention of not using form factors~\cite{PhysRevD.89.092005}.

For the \Zg process the existence of {\AGC}s would typically lead to an enhancement of photons with high transverse momentum~\cite{PhysRevD.30.1513, PhysRevD.57.2823,PhysRevD.47.4889}. The observed \ptg distribution is therefore used to extract limits on \ZZg and {\Zgg} {\AGC}s.

\begin{figure}[tbp]
\centering
\includegraphics[width=0.49\textwidth]{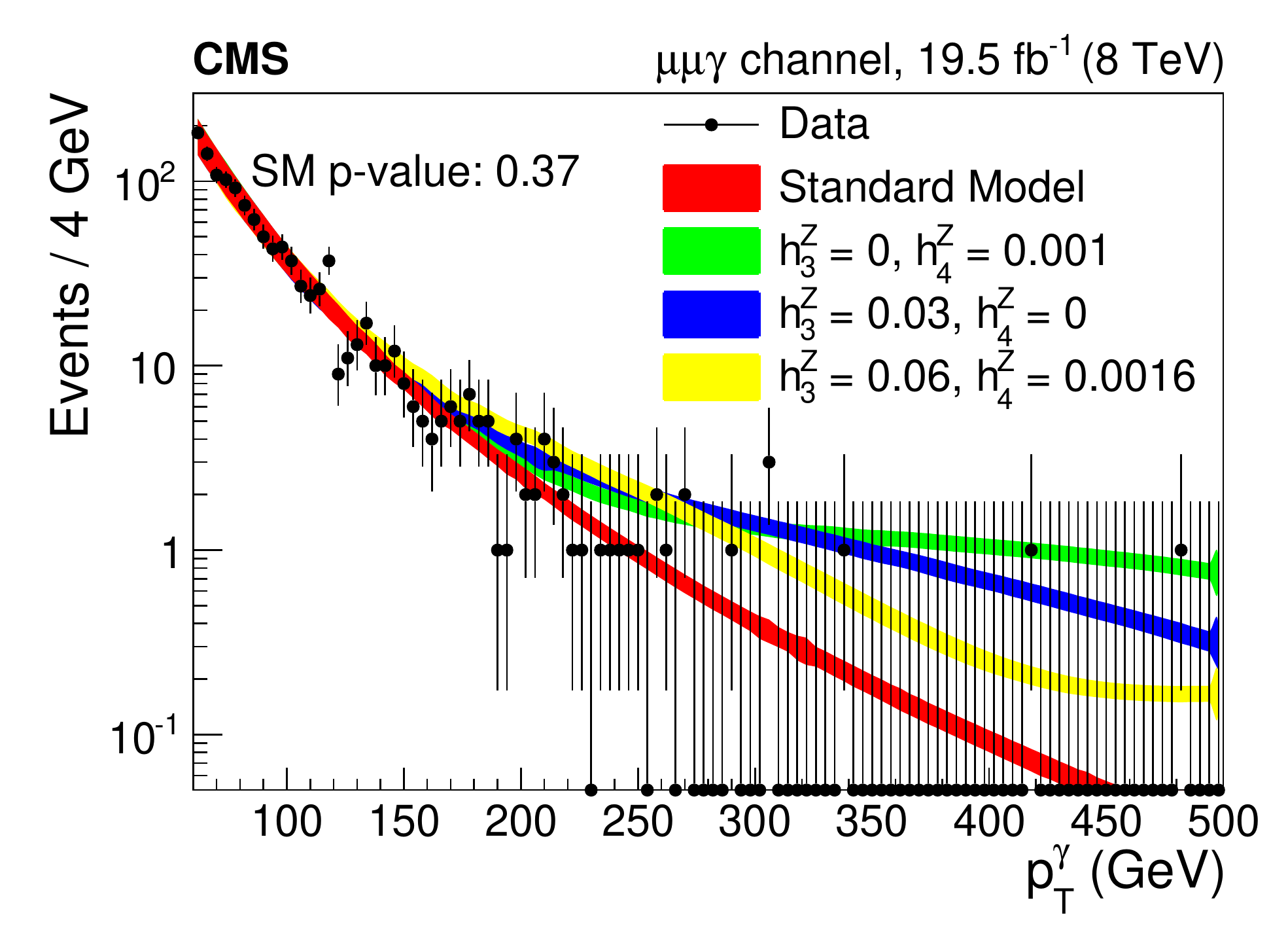}
\includegraphics[width=0.49\textwidth]{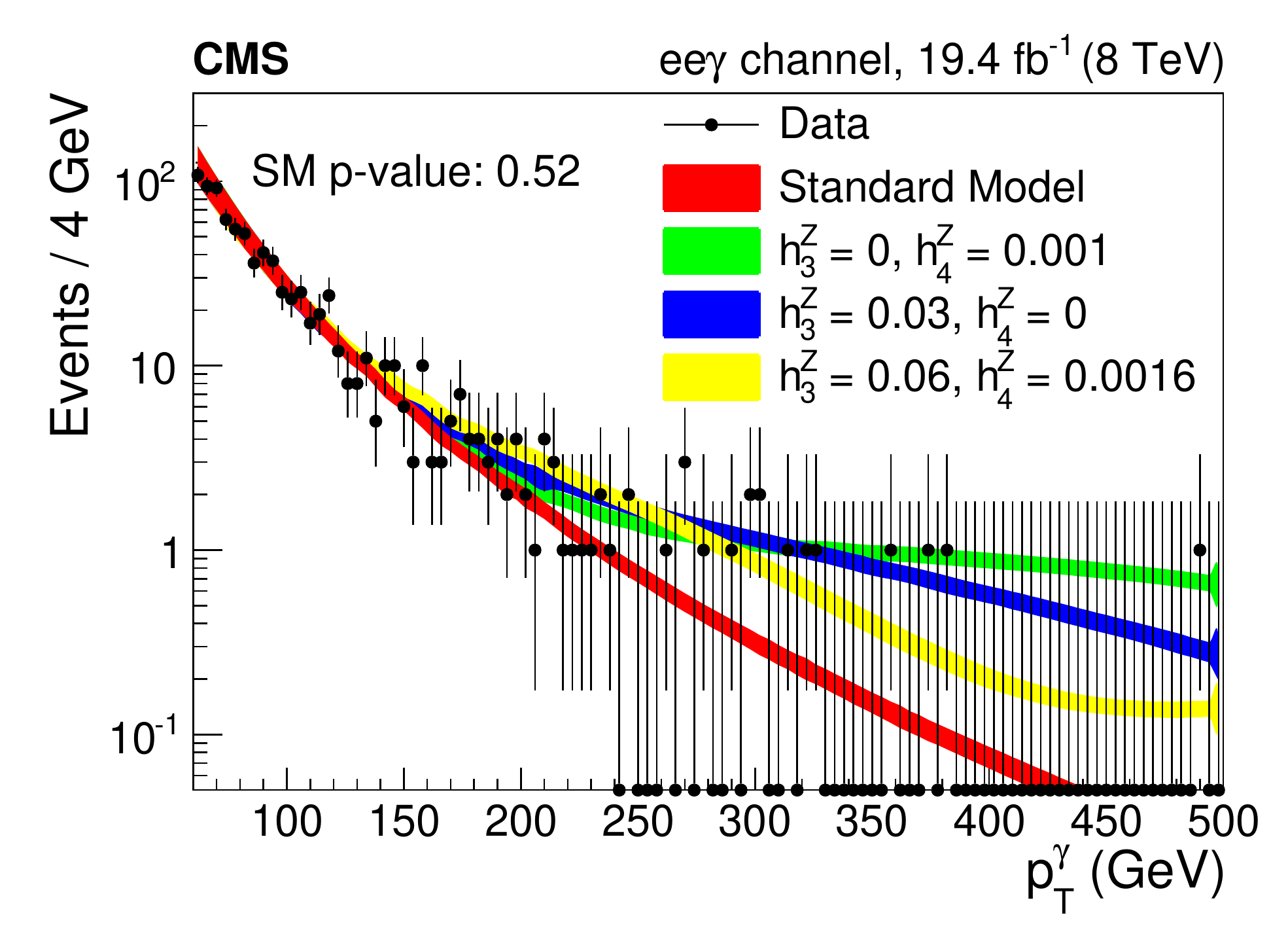}
\includegraphics[width=0.49\textwidth]{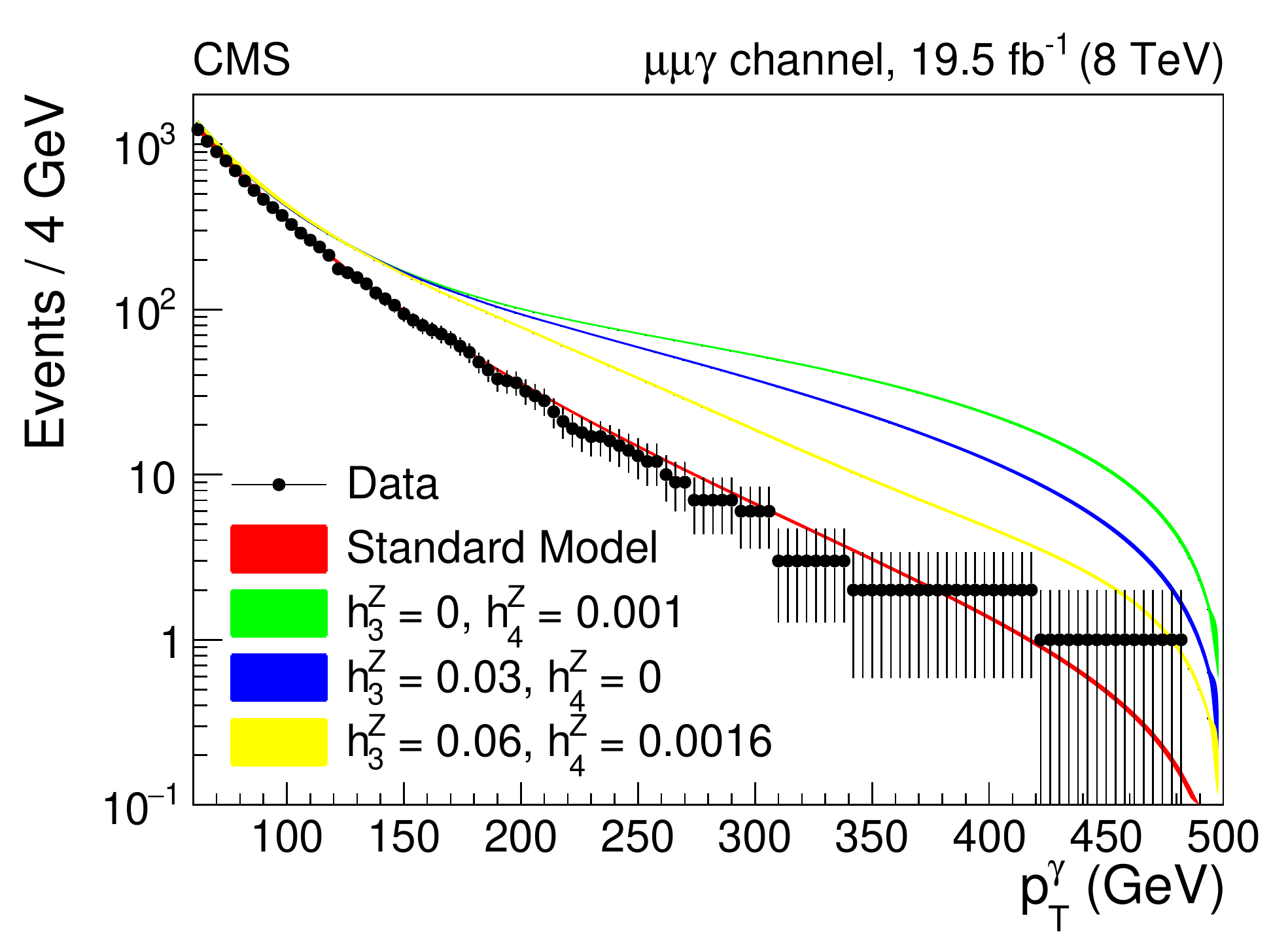}
\includegraphics[width=0.49\textwidth]{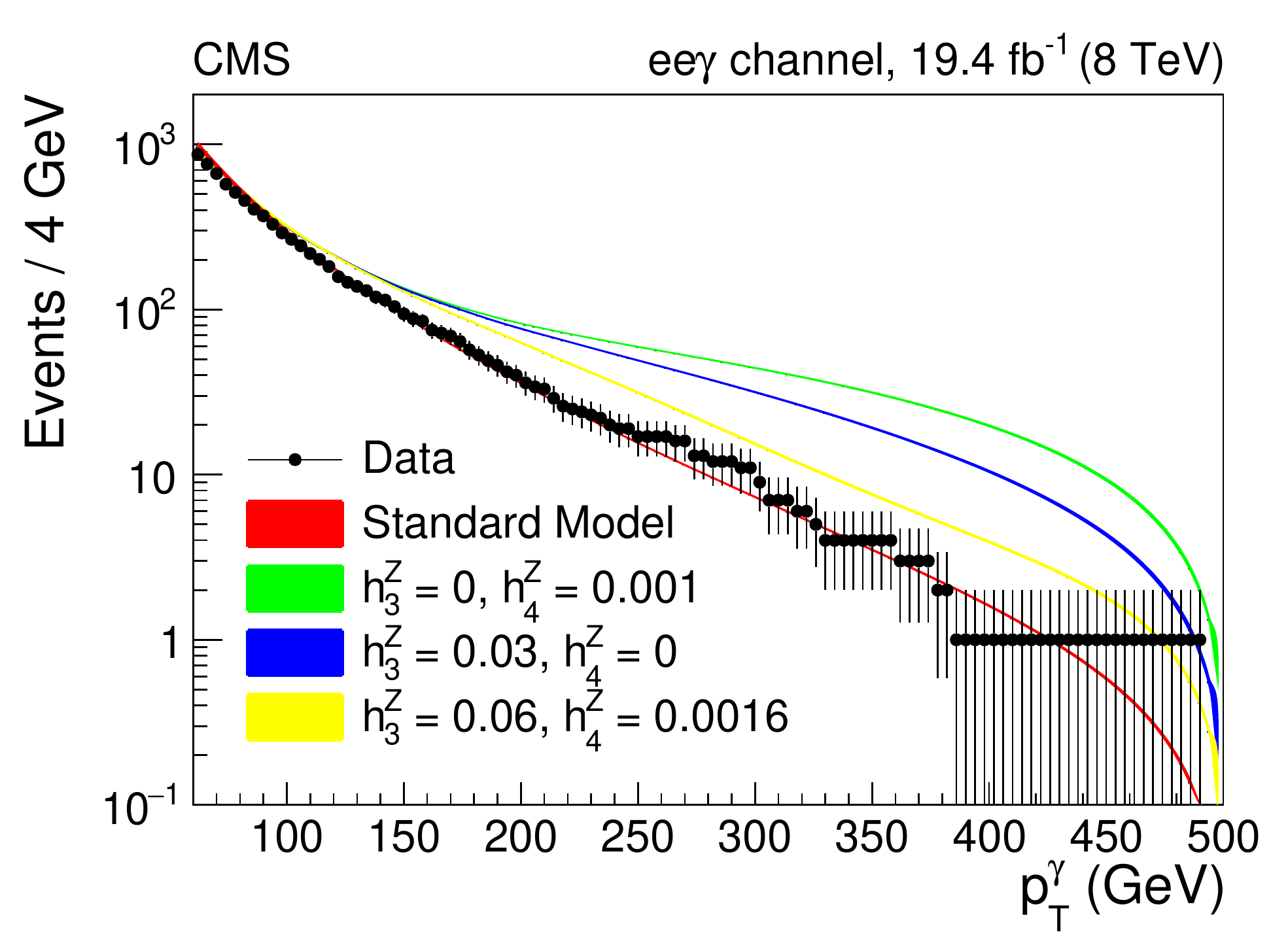}
\caption{Top: the \ptg distribution compared to predictions using various values for the {\AGC}s and the \SM. The observed $p$-values show that data are fully compatible with the \SM expectation (red). Bottom: corresponding cumulated distributions.}
\label{PMB1}
\end{figure}

The difference in the \ptg distributions between the \SM and {\AGC}s models is parameterized using the \MCFM (NLO) prediction. The NNLO \SM calculation is added to describe a complete \ptg distributions of an \AGC model. To obtain a $\ptg$ distribution that can be compared to the data, each simulated event is weighted by the lepton and photon efficiencies and the photon momentum is smeared according to the detector resolution. The irreducible background from the simulation and the background of nonprompt and misidentified photons, as obtained from the $\see$ template method, are added. In order to obtain a smooth background description, the background is parameterized as a sum of two exponential functions with parameters obtained from a fit to the observed background distribution. Figure~\ref{PMB1} shows a direct comparison between the $\ptg$ distribution in data and the expectations for various \AGC strengths. A theoretical uncertainty of 6--12\% is determined from PDF and scale variations. Experimental systematic uncertainties are 2\% in the dilepton efficiency, 2\% in the photon efficiency, 2.6\% in the luminosity measurement, and depending on $\ptg$ up to 8\% uncertainty in the background of nonprompt and misidentified photons obtained from the $\see$ template method.

\begin{figure}[tbp]
\centering
\includegraphics[width=0.7\textwidth]{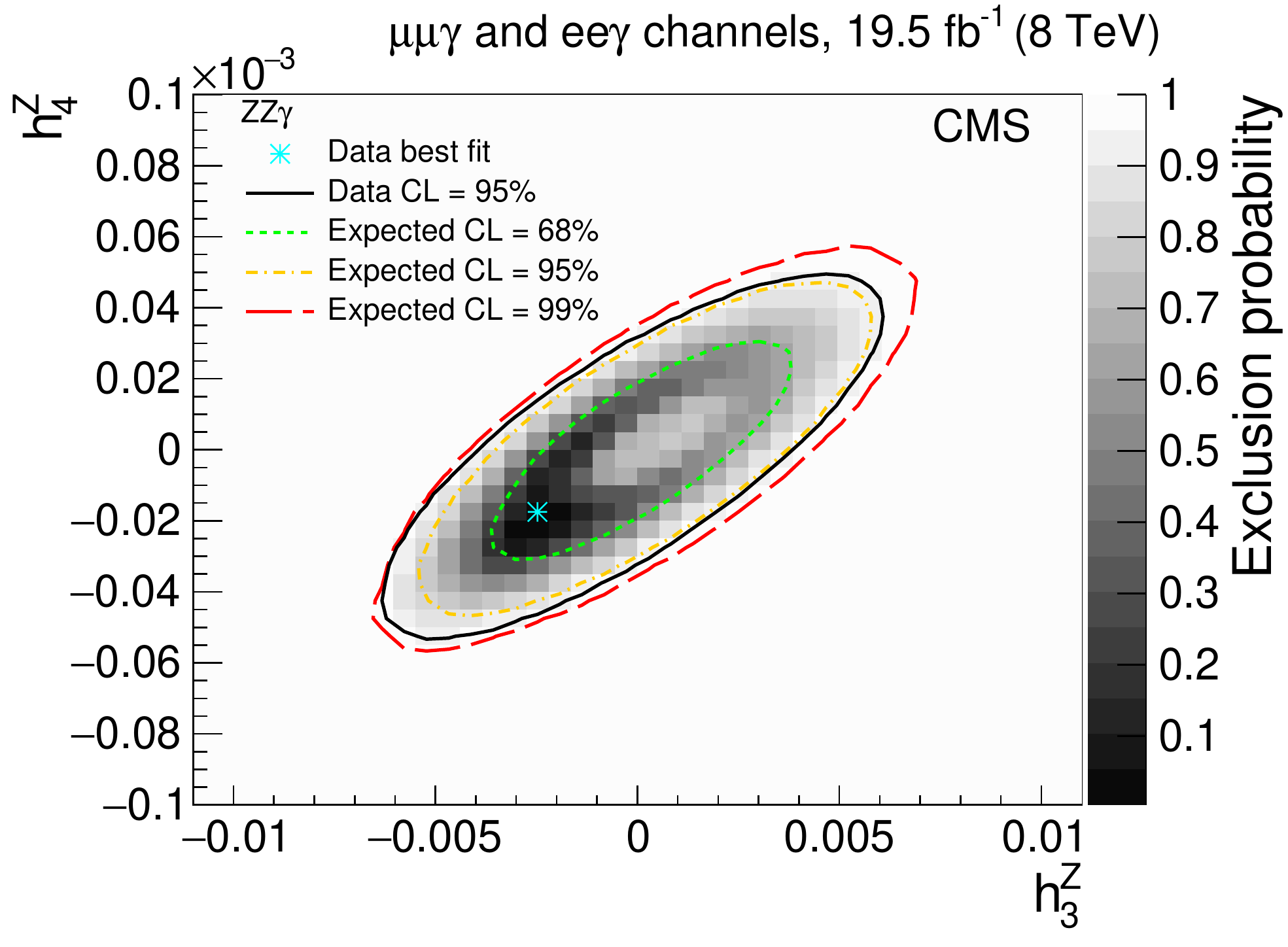}\\
\includegraphics[width=0.7\textwidth]{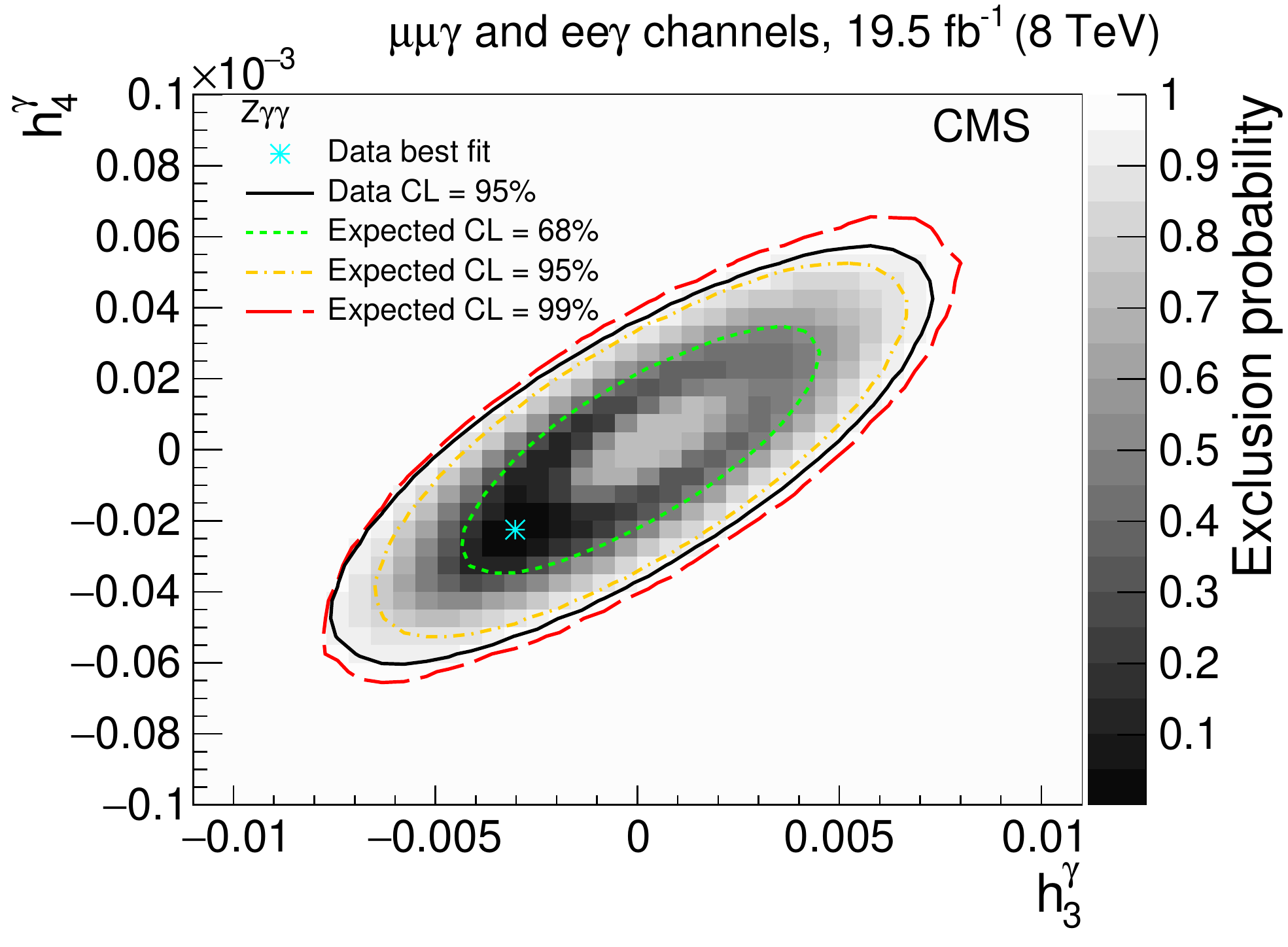}
\caption{Best fit of the combined muon and electron channels for models of anomalous \ZZg (top) and \Zgg (bottom) couplings. No form factor is used. The light blue star indicates the point of highest probability. The level of gray represents the exclusion probability and the black line corresponds to the 95\% CL limit. In addition, several expected contours from \SM simulation are shown.}
\label{PLC1}
\end{figure}

An unbinned profile likelihood ratio based on the \ptg distribution is used to find the best fitting {\AGC} model and its 95\% confidence level (CL) region. With the precision of the current measurement it is not possible to distinguish between the CP-even and CP-odd contributions. Therefore, only the CP-even parameters $h^V_3$ and $h^V_4$ are considered. The two-dimensional limits on $h_3^V$ and $h_4^V$ are shown in \FIG{PLC1}. The combination of the muon and electron channels takes into account that most of the systematic uncertainties are correlated with the exception of those related to the lepton reconstruction efficiencies. The one-dimensional 95\% CL regions, when only one of the {\AGC}s is nonzero, are
\begin{equation*}\begin{split}
-3.8\times10^{-3} &< h_3^Z < 3.7\times10^{-3}  \\
-3.1\times10^{-5} &< h_4^Z < 3.0\times10^{-5}  \\
-4.6\times10^{-3} &< h_3^\gamma < 4.6\times10^{-3}  \\
-3.6\times10^{-5} &< h_4^\gamma < 3.5\times10^{-5}.
\end{split}\end{equation*}

\section{Summary}
\label{sec:summary}
A study of \Zg production in pp collisions at 8\TeV using data collected with the CMS experiment in 2012, corresponding to an integrated luminosity of 19.5 \fbinv was presented. Decays of the \Z bosons into muons and electrons were used for the measurement of the differential \Zg cross section as a function of $\ptg$ for a phase space defined by the kinematic requirements on the final-state particles shown in~\TAB{TXS1}. In addition, the exclusive differential \Zg cross section for events with no accompanying central jets was presented. The inclusive and exclusive cross sections for $\ptg > 15$\GeV are measured to be:
\begin{equation*}\begin{split}
\sigma_\text{incl}&= 2063\pm 19\stat\pm 98\syst\pm 54\lum\fb,\\
\sigma_{\text{excl}}&= 1770\pm 18\stat\pm 115\syst\pm 46\lum\fb.
\end{split}\end{equation*}
Both values are compatible with the \SM expectations of $\sigma_\text{incl}^{\MCFM}= 2100\pm 120\fb$ ($\sigma_\text{incl}^{\mathrm{NNLO}}=2241\pm 22\fb$) and $\sigma_{\text{excl}}^{\MCFM}= 1800\pm 120\fb$, respectively. At high \ptg the inclusive measurement is well described by the NNLO calculation and also by the \SHERPA prediction including up to two additional partons at matrix element level, while a clear excess is observed with respect to the \MCFM (NLO) calculation. This emphasizes the importance of NNLO QCD corrections for this measurement. A similar excess is not observed for the exclusive measurement.

Limits on the strengths of anomalous \ZZg and \Zgg couplings have been extracted. The following one-dimensional limits at 95\% CL have been obtained
\begin{equation*}\begin{split}
-3.8\times10^{-3} &< h_3^Z < 3.7\times10^{-3}  \\
-3.1\times10^{-5} &< h_4^Z < 3.0\times10^{-5}  \\
-4.6\times10^{-3} &< h_3^\gamma < 4.6\times10^{-3}  \\
-3.6\times10^{-5} &< h_4^\gamma < 3.5\times10^{-5}.
\end{split}\end{equation*}
These limits are more stringent than previously published results on neutral {\AGC}s for the charged lepton decays of the \Z boson from LEP \cite{Delphizzg, Achard2004119, Abbiendi:2003va, LEPEWK}, Tevatron \cite{PhysRevLett.107.051802, Abazov:2011qp} and the LHC experiments \cite{Aad:2013izg, PhysRevD.89.092005}.
\begin{acknowledgments}
We thank Dirk Rathlev and Massimiliano Grazzini for providing us with the NNLO calculation of the cross section.

\hyphenation{Bundes-ministerium Forschungs-gemeinschaft Forschungs-zentren} We congratulate our colleagues in the CERN accelerator departments for the excellent performance of the LHC and thank the technical and administrative staffs at CERN and at other CMS institutes for their contributions to the success of the CMS effort. In addition, we gratefully acknowledge the computing centers and personnel of the Worldwide LHC Computing Grid for delivering so effectively the computing infrastructure essential to our analyses. Finally, we acknowledge the enduring support for the construction and operation of the LHC and the CMS detector provided by the following funding agencies: the Austrian Federal Ministry of Science, Research and Economy and the Austrian Science Fund; the Belgian Fonds de la Recherche Scientifique, and Fonds voor Wetenschappelijk Onderzoek; the Brazilian Funding Agencies (CNPq, CAPES, FAPERJ, and FAPESP); the Bulgarian Ministry of Education and Science; CERN; the Chinese Academy of Sciences, Ministry of Science and Technology, and National Natural Science Foundation of China; the Colombian Funding Agency (COLCIENCIAS); the Croatian Ministry of Science, Education and Sport, and the Croatian Science Foundation; the Research Promotion Foundation, Cyprus; the Ministry of Education and Research, Estonian Research Council via IUT23-4 and IUT23-6 and European Regional Development Fund, Estonia; the Academy of Finland, Finnish Ministry of Education and Culture, and Helsinki Institute of Physics; the Institut National de Physique Nucl\'eaire et de Physique des Particules~/~CNRS, and Commissariat \`a l'\'Energie Atomique et aux \'Energies Alternatives~/~CEA, France; the Bundesministerium f\"ur Bildung und Forschung, Deutsche Forschungsgemeinschaft, and Helmholtz-Gemeinschaft Deutscher Forschungszentren, Germany; the General Secretariat for Research and Technology, Greece; the National Scientific Research Foundation, and National Innovation Office, Hungary; the Department of Atomic Energy and the Department of Science and Technology, India; the Institute for Studies in Theoretical Physics and Mathematics, Iran; the Science Foundation, Ireland; the Istituto Nazionale di Fisica Nucleare, Italy; the Ministry of Science, ICT and Future Planning, and National Research Foundation (NRF), Republic of Korea; the Lithuanian Academy of Sciences; the Ministry of Education, and University of Malaya (Malaysia); the Mexican Funding Agencies (CINVESTAV, CONACYT, SEP, and UASLP-FAI); the Ministry of Business, Innovation and Employment, New Zealand; the Pakistan Atomic Energy Commission; the Ministry of Science and Higher Education and the National Science Centre, Poland; the Funda\c{c}\~ao para a Ci\^encia e a Tecnologia, Portugal; JINR, Dubna; the Ministry of Education and Science of the Russian Federation, the Federal Agency of Atomic Energy of the Russian Federation, Russian Academy of Sciences, and the Russian Foundation for Basic Research; the Ministry of Education, Science and Technological Development of Serbia; the Secretar\'{\i}a de Estado de Investigaci\'on, Desarrollo e Innovaci\'on and Programa Consolider-Ingenio 2010, Spain; the Swiss Funding Agencies (ETH Board, ETH Zurich, PSI, SNF, UniZH, Canton Zurich, and SER); the Ministry of Science and Technology, Taipei; the Thailand Center of Excellence in Physics, the Institute for the Promotion of Teaching Science and Technology of Thailand, Special Task Force for Activating Research and the National Science and Technology Development Agency of Thailand; the Scientific and Technical Research Council of Turkey, and Turkish Atomic Energy Authority; the National Academy of Sciences of Ukraine, and State Fund for Fundamental Researches, Ukraine; the Science and Technology Facilities Council, UK; the US Department of Energy, and the US National Science Foundation.

Individuals have received support from the Marie-Curie program and the European Research Council and EPLANET (European Union); the Leventis Foundation; the A. P. Sloan Foundation; the Alexander von Humboldt Foundation; the Belgian Federal Science Policy Office; the Fonds pour la Formation \`a la Recherche dans l'Industrie et dans l'Agriculture (FRIA-Belgium); the Agentschap voor Innovatie door Wetenschap en Technologie (IWT-Belgium); the Ministry of Education, Youth and Sports (MEYS) of the Czech Republic; the Council of Science and Industrial Research, India; the HOMING PLUS program of Foundation for Polish Science, cofinanced from European Union, Regional Development Fund; the Compagnia di San Paolo (Torino); the Consorzio per la Fisica (Trieste); MIUR project 20108T4XTM (Italy); the Thalis and Aristeia programs cofinanced by EU-ESF and the Greek NSRF; and the National Priorities Research Program by Qatar National Research Fund.
\end{acknowledgments}
\bibliography{auto_generated}

\cleardoublepage \appendix\section{The CMS Collaboration \label{app:collab}}\begin{sloppypar}\hyphenpenalty=5000\widowpenalty=500\clubpenalty=5000\textbf{Yerevan Physics Institute,  Yerevan,  Armenia}\\*[0pt]
V.~Khachatryan, A.M.~Sirunyan, A.~Tumasyan
\vskip\cmsinstskip
\textbf{Institut f\"{u}r Hochenergiephysik der OeAW,  Wien,  Austria}\\*[0pt]
W.~Adam, T.~Bergauer, M.~Dragicevic, J.~Er\"{o}, M.~Friedl, R.~Fr\"{u}hwirth\cmsAuthorMark{1}, V.M.~Ghete, C.~Hartl, N.~H\"{o}rmann, J.~Hrubec, M.~Jeitler\cmsAuthorMark{1}, W.~Kiesenhofer, V.~Kn\"{u}nz, M.~Krammer\cmsAuthorMark{1}, I.~Kr\"{a}tschmer, D.~Liko, I.~Mikulec, D.~Rabady\cmsAuthorMark{2}, B.~Rahbaran, H.~Rohringer, R.~Sch\"{o}fbeck, J.~Strauss, W.~Treberer-Treberspurg, W.~Waltenberger, C.-E.~Wulz\cmsAuthorMark{1}
% ...
\vskip\cmsinstskip
% ...
1:~~Also at Vienna University of Technology, Vienna, Austria\\
2:~~Also at CERN, European Organization for Nuclear Research, Geneva, Switzerland\\
% ...

\end{sloppypar}
\end{document}